\documentclass[preprint,aps,11pt]{revtex4}

\usepackage{graphicx,color}% Include figure files
\usepackage{dcolumn}% Align table columns on decimal point
\usepackage{bm}% bold math

\begin{document}

\title{\large \bf 
Neutrino Oscillations in the Precision  Era}  

\author{ M.~Bishai,  M.V.~Diwan, S. ~Kettell,  J. ~Stewart,   B. ~Viren, E. ~Worcester}
\affiliation{ Physics Department, Brookhaven National Laboratory, Upton, NY 11973}  

\author{Lisa Whitehead}
\affiliation{Department of Physics, University of Houston, Houston, TX 77204}

\date{\today} 
 
\baselineskip=12pt

\bigskip  

\begin{abstract}
%We examine experimental  approaches to search for hints of new physics in neutrino oscillations.  
With the discovery of a modest size for the mixing angle $\theta_{13} \sim 9^\circ$ by the Daya Bay collaboration at 
$>$5 sigma (\cite{dayabay})
the science of neutrino oscillations has shifted to 
explicit demonstration of CP violation and 
precision determination of the CP phase in the 3-flavor framework.  
Any additional contributions from new physics to the oscillation channel 
$\nu_\mu \to \nu_e$ could be uncovered by multiple constraints in the ($\theta_{13}, \delta_{CP}$) parameter
 space.  In long-baseline experiments 
such constraints will require examination of the oscillation strength at higher $L/E$ where the effects of 
CP violation will be large.  For the fixed baseline of 1300 km for the Long-Baseline Neutrino Experiment
 (LBNE, Fermilab to Homestake), it will be important 
to examine oscillations at low energies ($<1.5$ GeV) with good  statistics, low backgrounds, and excellent 
energy resolution.  The accelerator upgrades in the Project-X era have the potential to offer the beams of the 
needed intensity and quality for this advanced science program.  In this paper we examine the event rates for 
high intensity, low energy running of Project-X and the Fermilab Main Injector complex, 
and the precision in the ($\theta_{13}, \delta_{CP}$) space.
In this paper we have examined the baseline distance of 1300 km in detail, 
however we point out that much longer distances 
such as 2500 km should also be exmained with a beam from FNAL in light of the new understanding of the neutrino mixing.  

We find that the best way to obtain multiple constraints in the neutrino sector is  
to perform high statistics experiments
with low energy neutrino beams over long distances.  
For oscillation physics at low energies the charged current cross section is dominated by quasielastic scattering.
For the quasielastic final state any large 
detector capable of measuring single lepton final states is adequate. In the following we have used  
the water Cherenkov detector since it can be built to have the  target mass ($\sim$ 200kTon) needed
 to obtain the required statistical precision and, as a result of the work for LBNE, the performance of the 
detector at low energies is well established.

\end{abstract}

\maketitle

\section{Introduction}

The Daya Bay Reactor Neutrino Experiment has measured a non-zero value for 
the neutrino mixing angle $\theta_{13}$ with a significance of 5.2 standard
 deviations \cite{dayabay}. The value determined by using a rate-only analysis 
of the first data set from the experiment is 
$\sin^2 2 \theta_{13}= 0.092\pm 0.016 \pm 0.005$. The error currently is dominated 
by statistics.  

This determination of a modest size 
  for the third mixing angle ($\theta_{13} \sim 9^\circ$) in 
the neutrino sector allows us to plan for the future program of precision neutrino physics. 
With this value for the third mixing angle, 
 a rich new system has  become 
available for the physics of neutrino oscillations, and leptonic CP violation. 

Now that the parameters of neutrino mixing have been  measured it will be necessary to exploit the 
new system to find indications of departures from the Standard Model with massive neutrinos ($\nu$SM). These 
departures could come from additional low energy interactions or unexpected mixing. It is not possible to 
predict the exact nature of new physics or investigate every model for new neutrino physics that is now 
being discussed in the theoretical literature.  Nevertheless, it is possible to sketch
 some generic attributes regarding 
 the design of a new experiment that is  likely to yield indications of new physics.  
 These necessary attributes would be 
based on physics that is  known, physics that has already been excluded and reasonable extrapolations of 
scientific strategies that have  worked in the past.

\begin{figure} 
\includegraphics[angle=0,width=0.49\textwidth]{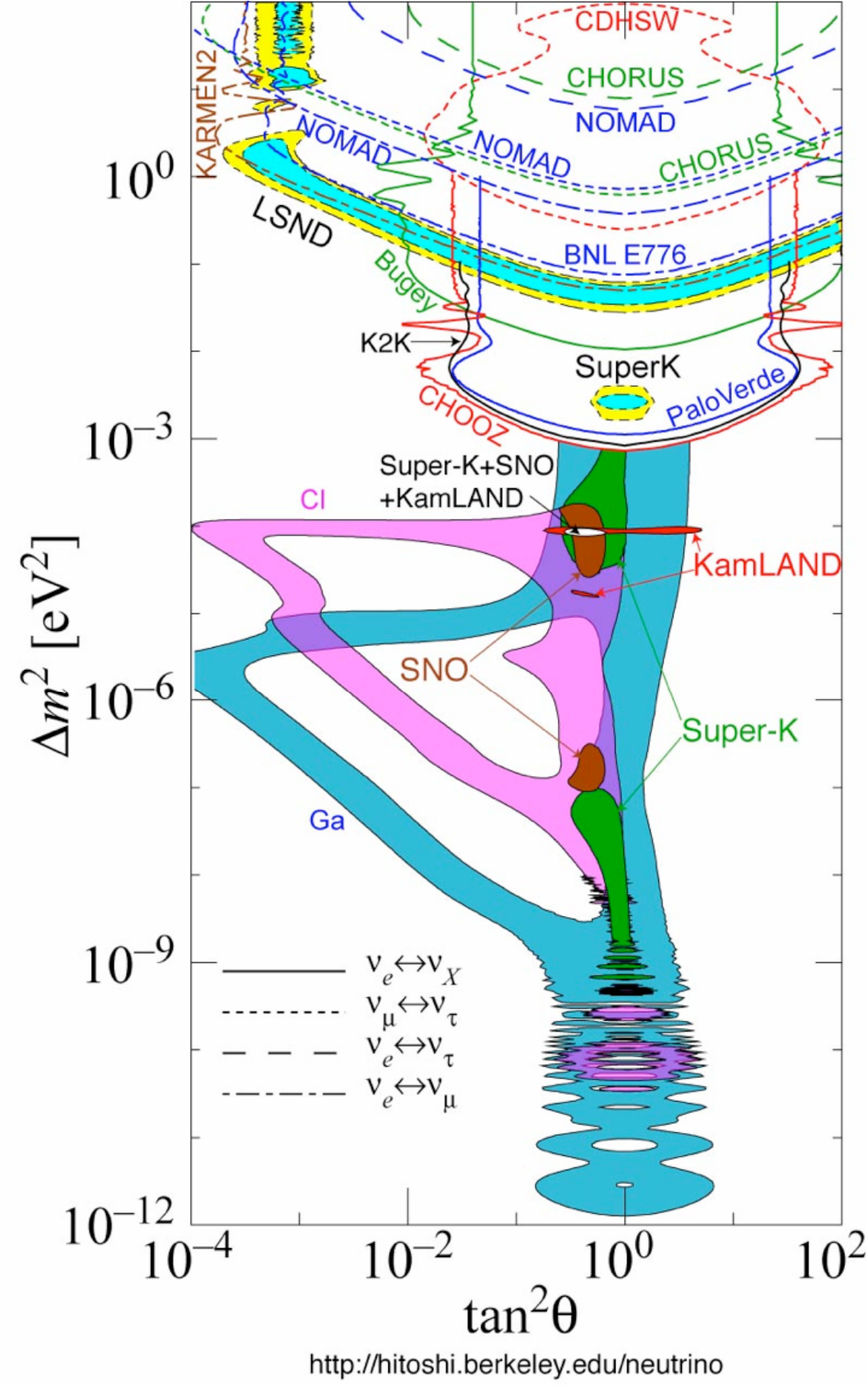}
\caption{Compilation of neutrino oscillation limits and results. The filled areas are positive
 indications and lines correspond to 
90\% confidence level limits unless otherwise indicated. \cite{hitoshi} }
\label{hitoshi}
\end{figure}

An almost complete history of neutrino oscillation experiments (except for the Daya Bay experiment)  
can be found  in figure \ref{hitoshi}.  
There are some important lessons to be learned in this figure. 
Experiments designed for high energy operation probe  mixing in the region of $\Delta m^2 > 1 eV^2$.
Many such experiments were carried out with sensitivity to smaller and smaller 
mixing angles with limited success.  Eventually mixing was found to be large and the correct mass 
range to be much smaller than 1 $\rm eV^2$. 
Previous results as well as the direct limits on neutrino mass from 
tritium beta decay ($m_{\nu_e} < 2$ eV at 95\% C.L.) 
 and astrophysics ($\sum_i m_{\nu_i} < $0.4 -- 1 eV) 
have put severe limits on physics at $L/E < 1$ km/GeV \cite{ref2}. 
Furthermore, any remaining issues such as the LSND/MiniBoone/reactor anomalies   
are best resolved by short baseline accelerator experiments. 
  As we examine the oscillation phenomena in greater detail we find that large interesting effects 
are always at high $L/E$. 
Moreover, most of the precision measurements in neutrino physics have been
performed at low energies.  
Experiments that have yielded the best results are 
 at high $L/E > 500$ km/GeV, or for any fixed distance, the lowest 
energy neutrinos have provided the best science.

In the following, we first calculate the oscillation probability for $\nu_\mu \to \nu_e$ as a function of energy 
using the best values for the oscillation parameters.  We will then calculate the energy spectrum of neutrinos 
that could be produced by various beams from Fermilab. 
Finally we calculate the precision in the ($\theta_{13}, \delta_{CP}$)
space for an example running condition, and comment on constraints on other parameters 
that could result from the data. 

\begin{figure} 
\mbox{\includegraphics[angle=0,width=0.45\textwidth]{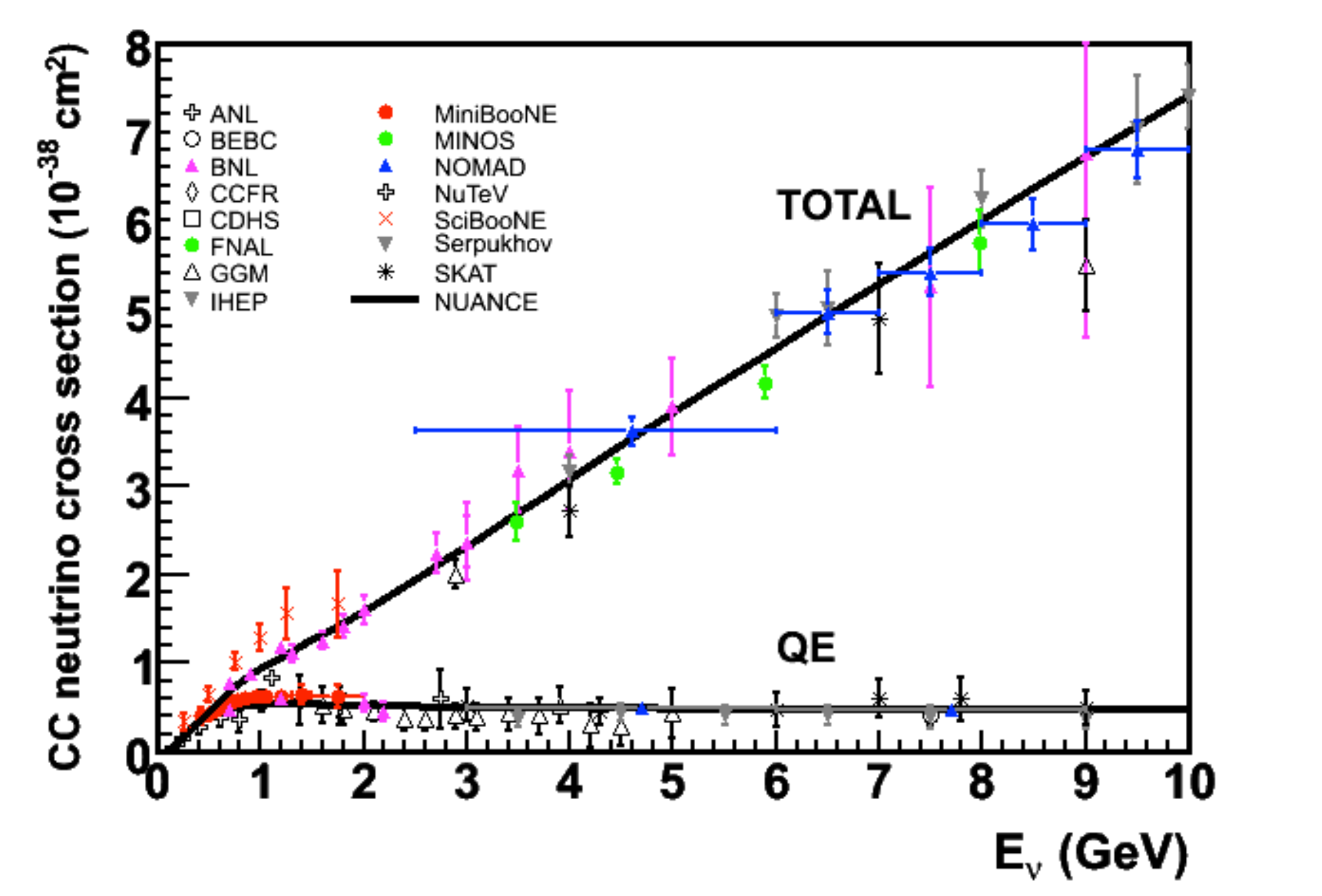}
\includegraphics[angle=0,width=0.45\textwidth]{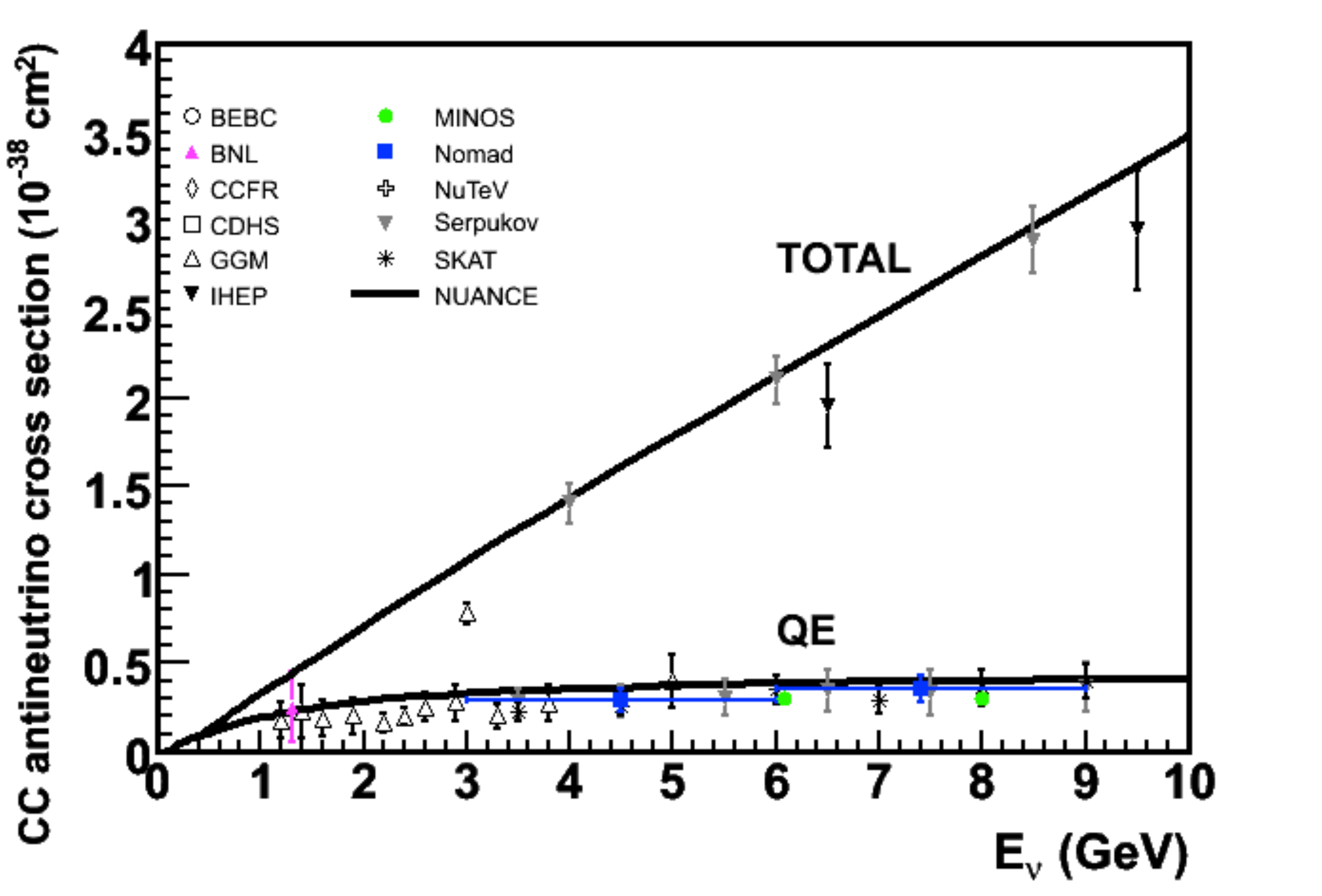}}
\caption{Charged current neutrino total cross section as a function of energy for neutrinos (left) and 
antineutrinos (right). The cross section of quasielastic interactions is also displayed. The error bars are 
the data and the lines are the current best model.  It should be noted that below 1.5 GeV the total cross section is 
dominated by quasielastic interactions.\cite{zeller}   }
\label{crssec} 
\end{figure}

For oscillation physics at low energies the charged current cross section 
is dominated by quasielastic scattering as 
shown in figure \ref{crssec}.
For the quasielastic final state any large detector capable of measuring single lepton final states is adequate. 
The water Cherenkov detector  can achieve the target mass needed to obtain the 
statistical precision at low energies with high efficiency.  In the following we will therefore assume
 a 200 kTon fiducial water Cherenkov detector for the  target mass. Furthermore,  
as a result of the work for LBNE the performance of the 
detector at low energies is well established.    

\section{$\nu_\mu \to \nu_e$ Oscillations} 

The oscillation mode with the richest scientific content is the appearance mode 
$\nu_\mu \to \nu_e$ (or $\nu_e \to \nu_\mu$) and 
its antineutrino counterpart. The oscillation strength for this mode as a function of distance and energy 
contains contributions from all parameters of the neutrino mixing matrix and all three neutrino mass eigenstates. 

Assuming a constant matter density, 
which is a good approximation for the baselines in question,
the oscillation of $\nu_{\mu} \rightarrow \nu_e$ in the Earth 
for 3-generation mixing is described
approximately by the following equation \cite{freund}

\begin{eqnarray}
P(\nu_{\mu} \rightarrow \nu_e) &\approx&
\sin^2 \theta_{23} {\sin^2 2 \theta_{13}\over (\hat{A}-1)^2}\sin^2((\hat{A}-1)\Delta)  \nonumber\\ &&
+\alpha{\sin\delta_{CP}\cos\theta_{13}\sin 2 \theta_{12} \sin 2
\theta_{13}\sin 2 \theta_{23}\over
\hat{A}(1-\hat{A})} \sin(\Delta)\sin(\hat{A}\Delta)\sin((1-\hat{A})\Delta) \nonumber\\ &&
+\alpha{\cos\delta_{CP}\cos\theta_{13}\sin 2 \theta_{12} \sin 2
\theta_{13}\sin 2 \theta_{23}\over
\hat{A}(1-\hat{A})} \cos(\Delta)\sin(\hat{A}\Delta)\sin((1-\hat{A})\Delta) \nonumber\\ &&
+\alpha^2 {\cos^2\theta_{23}\sin^2 2 \theta_{12}\over
\hat{A}^2}\sin^2(\hat{A}\Delta) \nonumber \\
\label{qe1}
\end{eqnarray}

where $\alpha=\Delta m^2_{21}/\Delta m^2_{31}$, $\Delta = \Delta
m^2_{31} L/4E$, $\hat{A}=2 V E/\Delta m^2_{31}$,
$V=\sqrt{2} G_F n_e$. $n_e$ is the density of electrons in the Earth. 
Recall that $\Delta m^2_{31} = \Delta m^2_{32}+\Delta m^2_{21}$. 
Also notice that $\hat{A}\Delta = L G_{F} n_e/\sqrt{2}$ is sensitive 
to the sign of $\Delta m^2_{31}$. 
For antineutrinos, the second term in Equation \ref{qe1}  
has the opposite sign, and the matter potential also has the opposite sign. The second term  is
proportional to the following CP violating quantity.

\begin{equation}
J_{CP} \equiv \sin\theta_{12} \sin\theta_{23} \sin\theta_{13} \cos\theta_{12} 
\cos\theta_{23} \cos^2\theta_{13} \sin \delta_{CP}
\label{eq2}
\end{equation}

Equation \ref{qe1}
is an  expansion in powers of $\alpha$. This approximate formula is
useful for understanding important features of the appearance probability:
1) the first three terms in the equation control the matter induced enhancement for 
normal mass ordering ($m_1< m_2< m_3$) 
or suppression for the reversed mass ordering ($m_3< m_1< m_2$) of the oscillation 
probability above 3 GeV; 
2) the second and third terms control the sensitivity to
CP in the $\sim$ 1 GeV range; and 3) the last term controls
the sensitivity to $\Delta m^2_{21}$ at low energies. The first term (last term) is also 
proportional $\sin^2 \theta_{23}$ ($\cos^2 \theta_{23}$), and therefore is sensitive to the issue of maximum 
mixing in  $\theta_{23}= \pi/4$.
As previously explained \cite{ref3}, measurement of the spectrum of oscillated $\nu_e$ events 
will allow us access to all of these parameters in a single experiment with good control of 
systematics. 

\begin{figure} 
\includegraphics[angle=0,width=0.9\textwidth]{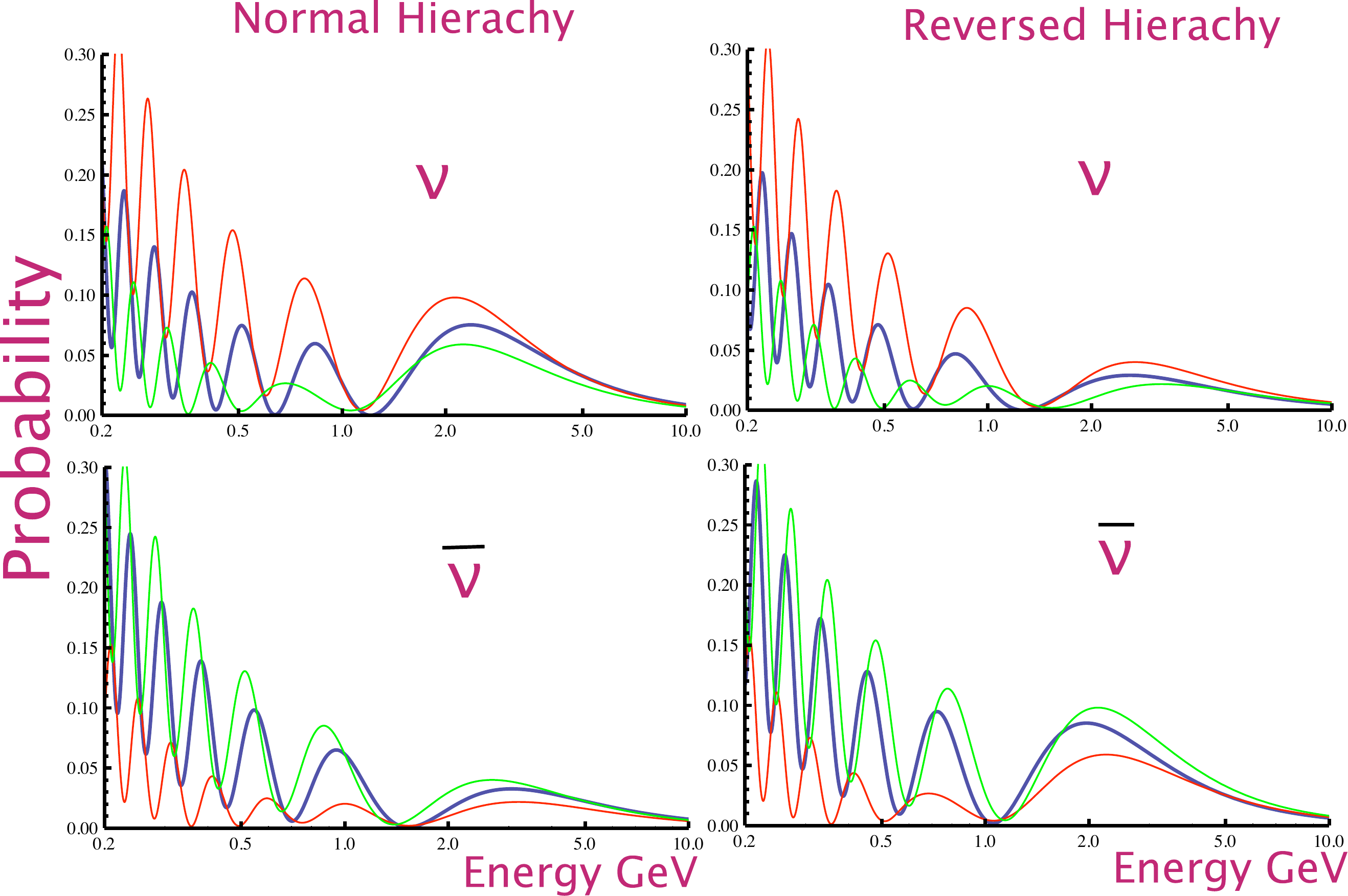}
\caption{ Appearance probability for $\nu_\mu \to \nu_e$  as a function of energy at a distance of 
1300 km. The top plots are for neutrinos and bottom plots are for antineutrinos.  The left side plots are
for normal mass ordering and right hand side are for reversed mass ordering.  The parameters used for these 
plots are $\Delta m^2_{32} = 0.0025 eV^2$, $\Delta m^2_{12} = 7.6 \times 10^-5 eV^2$, $\theta_{23} = \pi/4$,
$\theta_{12} = 34^\circ$, $\theta_{13} = 9.2^\circ$. The blue curve in all cases is for $\delta_{CP} = 0$ and the red
and green curves are for $\delta_{CP} = \pi/2$ and $-\pi/2$, respectively. 
} 
\label{oscplots}
\end{figure}

The oscillation probability for $\nu_\mu \to \nu_e$ for both neutrino and antineutrino modes and for 
normal and reversed mass ordering is plotted in figure \ref{oscplots} using close to the 
current best known parameters \cite{fogli} and rounding the Daya Bay 
measurement to the value  $\sin^2 2 \theta_{13} = 0.1$. 
It is easy to see that the oscillation probability 
will be in the 5-10 \% range at the first oscillation maximum between 2 - 3 GeV, but at lower energies both the 
size of the probability and the effect of the CP phase will be  dramatic. The lower energy oscillation effect 
is also relatively less affected by the mass hierarchy.  A measurement of these probabilities across the 
energy range will certainly result in precise new information about the mechanism of neutrino oscillations.

\section{Event rates for LBNE for various Project-X  beam conditions}

The relevant energy range of oscillations over 1300 km can be seen from figure 1 to be from 0.2 GeV to 4 GeV. 
Any flux above 4 GeV is  not going to contribute much to physics, and events  below 0.2 GeV will 
most likely have poor resolution.  Neutrinos in the 0.2 to 4 GeV energy range are produced by pions of 
energy $\sim 0.5-10$ GeV.  We shall assume that the proton energy available from  Project-X and related upgrades 
will be in the 8 to 120 GeV range.  
 The production spectrum of neutrinos in the laboratory frame is affected by 
the proton energy, the focusing efficiency of the target/horn system, and finally the kinematics of pion decay. 
In the center of mass of proton collisions, pion production is distributed in an approximate normal distribution 
as a function of rapidity and increases only logarithmically as a function of $\sqrt{s}\propto \sqrt{E_{protons}}$.  
In the laboratory frame this 
distribution of pions is boosted by a  factor $\gamma_{cm} \sim \sqrt{E_{proton}}$. 
For higher proton energies, there are higher number of pions produced at high rapidity in the center of mass,  
and they are boosted to 
higher energies in the laboratory frame. Both effects are such that in our proton and pion
energy range, the production of pions at any given energy  is roughly proportional to  
the energy of the protons and, obviously, the total current of protons, or the total proton beam power.  
If we assume that the target/horn system can be tuned to focus pions efficiently at any energy above 0.5 GeV, 
the neutrino yield is further affected by the kinematics of pion decay. The flux of neutrinos from pion decay in the
forward region is proportional to $\gamma^2_\pi \propto E^2_\pi$.  The two kinematic effects -- the production of pions 
as a function of proton energy and the pion decay kinematics -- are such that the only way to boost the yield of 
neutrinos at low energies is with high proton beam power at low energies.  

\begin{figure} 
\includegraphics[angle=0,width=0.9\textwidth]{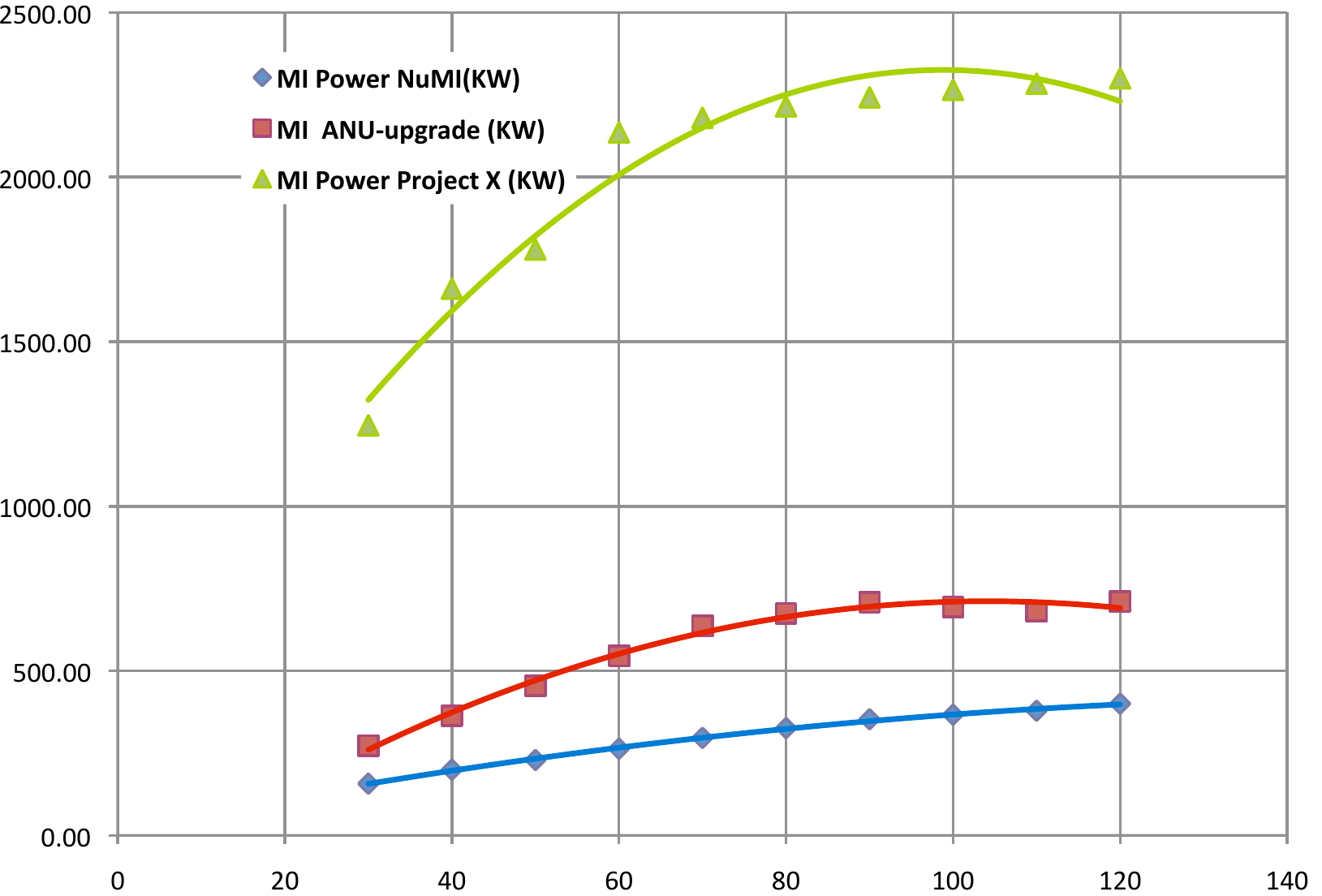}
\caption{ Proton beam power as a function of proton energy from the Fermilab 
Main Injector. Shown are current capabilities labeled as NuMI. The recently funded 
upgrades (labeled as ANU) will increase the power to 550 kW at 60 GeV or 700 kW at
120 GeV.  Project-X as currently conceived will allow beam power of 2 MW at 60 GeV and 
2.3 MW at 120 GeV. \cite{stevebob}  }
\label{proxfig} 
\end{figure}

In table \ref{protab} we have made a list of beam conditions that could be possible from  Project-X 
and the Project-X upgrade at 8 GeV.   Figure \ref{proxfig} shows the beam power available from the Main Injector 
as a function of energy.  With Project-X the beam power from
 the Main Injector can be maintained  at or above 2 MW over the range 60 - 120 GeV. 
This is because the decrease in energy can be (mostly) compensated by increasing the repetition rate. This trend 
continues as the energy decreases but at some point  it is limited by the number of  protons coming from 
the LINAC. The power achievable at 30 GeV would be  $\sim$1.3 MW for the Project-X Reference Design.
The Main Injector requires 270 kW of incident 8 GeV beam power at 8 GeV to produce $\sim$ 2 MW at 
60 GeV. With the additional upgrade to the 8 GeV pulsed LINAC, the 8 GeV power level could be increased to 
$\sim$ 4 MW. In such a scenario, the Fermilab accelerator complex could produce multi-MW power at both 
60 GeV and 8 GeV simultaneously. The duty factor for any Main Injector operation would continue to remain 
small in the single turn extraction mode, however the duty factor at 8 GeV will be $\sim 5-10 \%$ unless 
a ring is deployed to compress the beam further.  The poor duty factor is not problematic as long as the far detector 
is deployed at depth
in order to reduce backgrounds associated with cosmic rays. 
The duty factor might be more important for the operation of the target/horn system. We will assume 
that a ring might be deployed to produce an appropriately short duty factor $\sim 10^{-3}$. 

For our  calculations of event rates and neutrino spectra, we will use operation at 60 GeV at 
2 MW  and operation at 8 GeV at 3 MW.   We have reduced the power assumption for 8 GeV running from 4 MW to 3 MW 
because some of the current (270 kW) will be needed for producing the simultaneous 60 GeV beam, and the rest 
might be needed for other experiments. 
We have calculated the beam spectra using a GEANT4 simulation of 
the LBNE  beamline with magnetic horns  with a current of 
 250 kAmps, 2 meter diameter decay tunnel with a length of 280 meters.  
The LBNE beamline was designed for high energy operation. There is currently no  design to transport 
8 GeV protons to the LBNE target. The 8 GeV beam will require either another beamline or substantial 
modifications to the current beamline design.   
Here we will not investigate these important technical issues regarding the beamline, but 
we will argue  that  simultaneous operation at 
8 and 60 GeV  is quite compelling and should be investigated. 
In the following  we  focus on the 
neutrino spectra and event rates.

%Slight clarifications to Bob's response:   We are able to maintain the beam power from
% the Main Injector at or above 2 MW over the range 60 - 120 GeV. This is because the 
%decrease in energy can be (mostly) compensated by increasing the rep rate. This trend 
%continues as the energy decreases but at some point we run out of protons coming from 
%the linac. The power achievable at 30 GeV would be more like 1.3 MW for the PX Reference Design.
%The duty factor from the Main Injector is the pulse length, about 9.5 usec (a little
% less than Bob said), divided by the cycle time - currently specified as 1.3 sec at 
%120 GeV and 0.7 sec at 60 GeV. So it's in the range 7E-6 to 1.3E-5. 
%The Main Injector requires (8-GeV/60-GeV) * 2020 kW = 270 kW of
%incident 8 GeV beam power to generate 2020 kW of beam power at 60 GeV.
%Hence the available beam power at 8 GeV in the beyond-baseline scenario
%would increase by 7% to 4270 kW in the case on no Main Injector running.
%We do not have a detailed design yet on how the 8 GeV beam power would be
%transported from the 8 GeV pulsed linac to the LBNE beamline.  In all cases
%an accumulator/compressor ring is required to transform the  relatively high
%duty factor 8 GeV linac beam into low duty factor (<10**-3) beam suitable
%for LBNE.   The existing 8 GeV Recycler Ring could serve as a compressor
%ring, although we are also considering other solutions for the
%accumulator/compressor ring.   

\begin{table} 
\begin{tabular}{|l|c|c|c|c|}
\hline 
Accelerator &  Energy & Current & Duty & Power \\
Stage &  & & Factor & Available  \\ 
\hline  
CW  & 3 GeV & 1 mA & CW & 3000 \\  
LINAC  &    &      &    & kW    \\ 
\hline 
Pulsed &  8 GeV & 43 $\mu$A & 4.33ms/0.1sec & 350 \\ 
LINAC  &      &           &    &  kW  \\  
\hline  
8 GeV   &  8 GeV & 500 $\mu$A & 6.67ms/0.066sec &  4000 \\  
Upgrade  &      &             &       &   kW  \\ 
\hline  
Main   &  60 GeV & 35 $\mu$A &  9.5$\mu$S/0.7sec  &  2100  \\  
Injector  &   &  &  &   kW  \\  
\hline
Main &   120 GeV &  19 $\mu$A & 9.5$\mu$S/1.3sec & 2300   \\
Injector  &    &              &       & kW \\ 
\hline  
\end{tabular} 
\caption{  Beam conditions and power possible during the Project-X phase.  
An accumulator ring at 8 GeV could be used to improve the duty factor.  
\cite{stevebob}} 
\label{protab} 
\end{table}

\begin{figure} 
\mbox{\includegraphics[angle=0,width=0.45\textwidth]{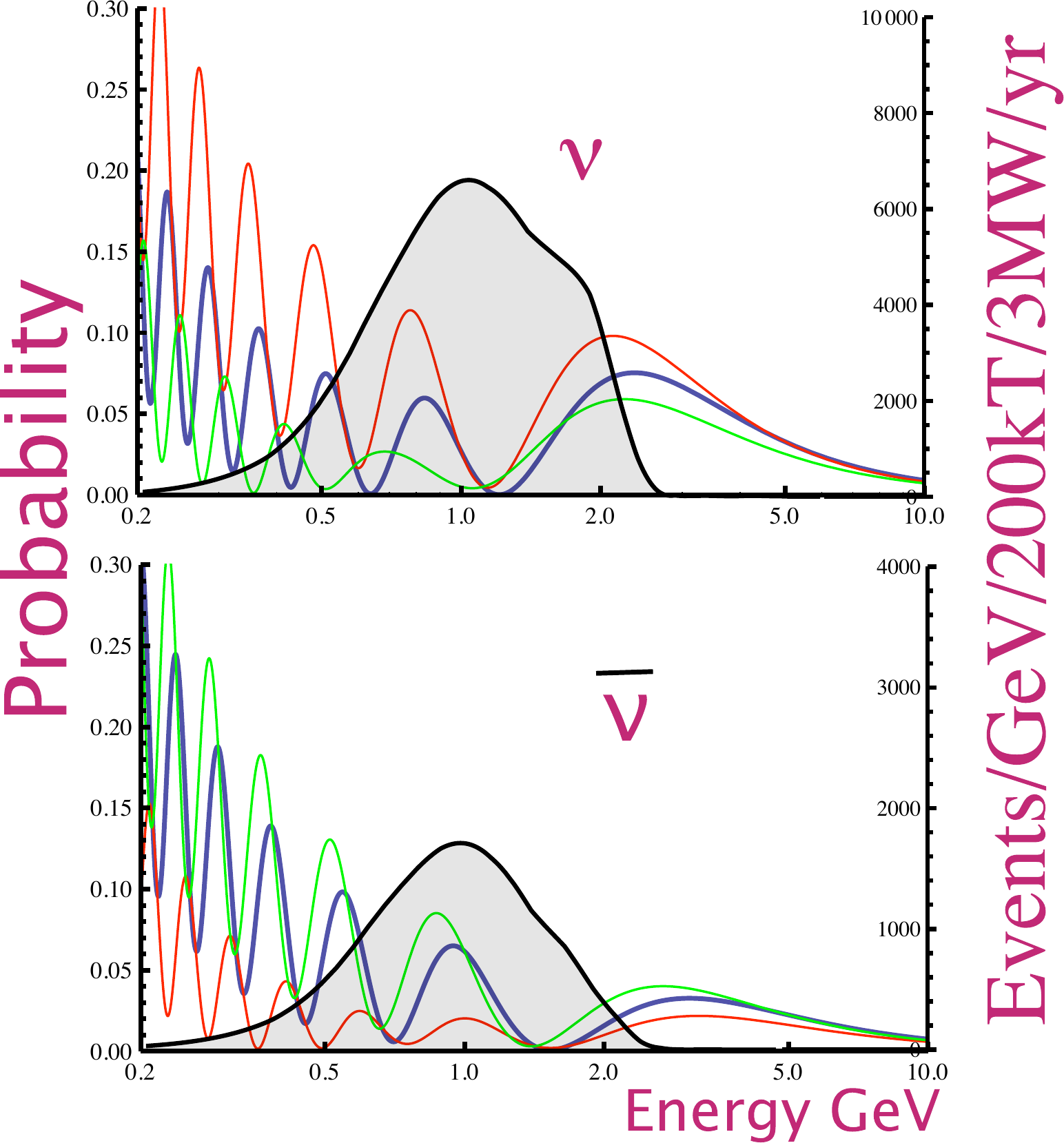}
\includegraphics[angle=0,width=0.45\textwidth]{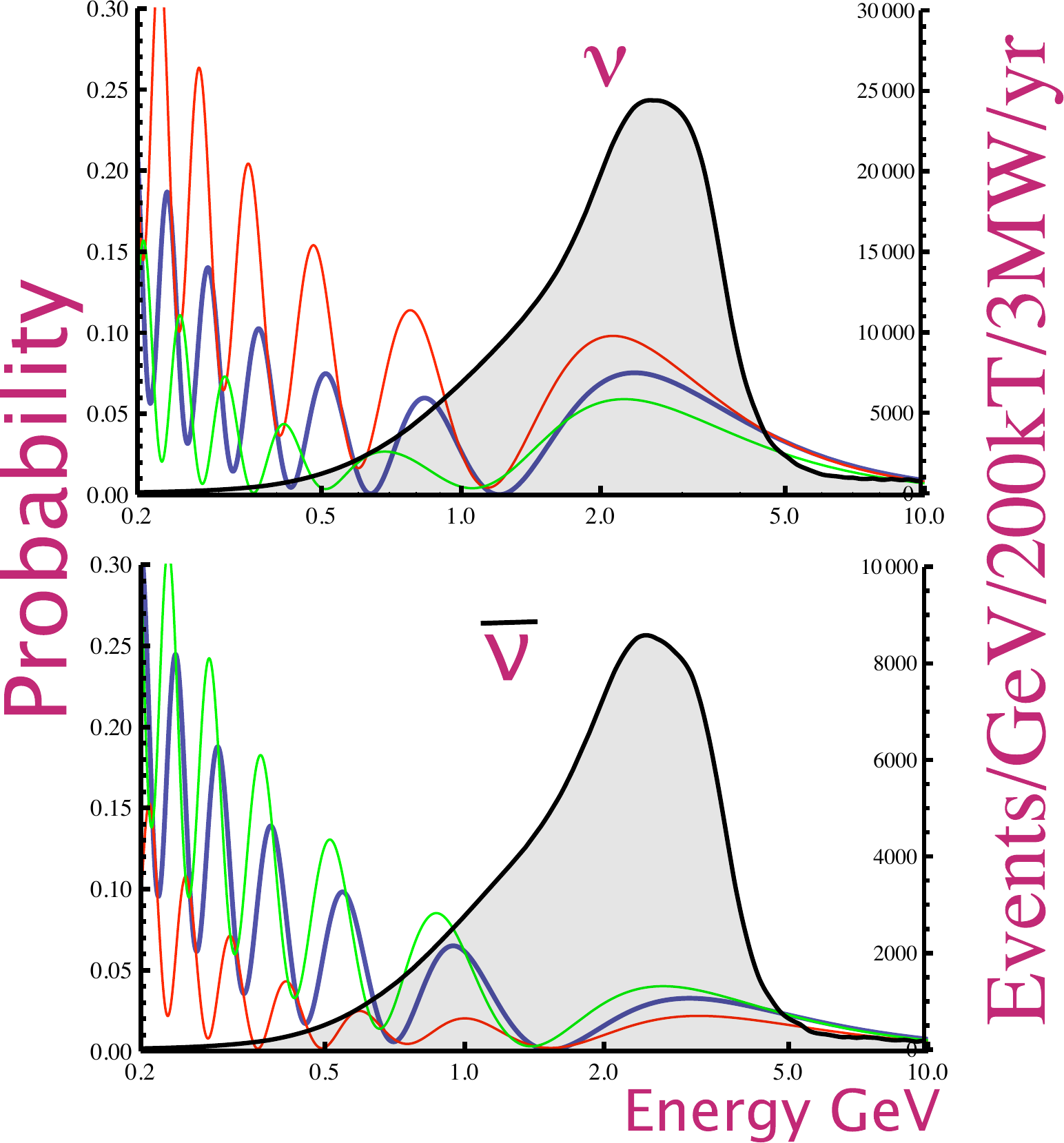}}
\caption{
Spectra of event rates as a function of energy for 8 GeV (left) and 60 GeV (right) proton beams 
From Fermilab. The spectra are superimposed on the expected oscillation probability for normal 
hierarchy.  Spectra are for 
the total charged current cross-section for muon neutrino (top) and antineutrinos (bottom). The beam is from 
Fermilab to Homestake over a distance of 1300 km; the intensity for the 8 GeV beam is assumed to be 
3 MW and for 60 GeV it is 2 MW. The detector size is 200 kTon fiducial mass.   
}
\label{spec} 
\end{figure}

The muon neutrino and antineutrino spectra (without oscillations)  for 8 GeV and 60 GeV beams are shown in 
figure \ref{spec} superimposed on the expected $\nu_\mu \to \nu_e$ oscillation probability. 
The event rate is calculated for the total muon neutrino (and antineutrino) cross section shown in 
figure \ref{crssec}. The two different energy spectra are shown to complement each other. The 8 GeV 
spectrum covers the low energy region where large CP phase effects exist while the 60 GeV spectrum 
covers the higher energy region where the matter effects will dominate. It should be remarked that the 
60 GeV beam also has similar numbers of  events at low energies as the low energy beam, but the low energy beam
 is expected to have somewhat more rate and less backgrounds due to event mis-reconstruction.  
The beam contamination in these beams is  shown in table \ref{ratetab} where the 
total event rate is tabulated for each component of the beam. The event rate after 
$\nu_\mu$ disappearance is also shown for the muon neutrino component. A few comments are in order regarding 
this table: 

\begin{itemize}
\item  The event rates have been calculated for the total cross section as in figure 
\ref{crssec}. In the next section we will use the tabulated 
water Cherenkov detector performance 
for extracting electron neutrino events  and associated backgrounds. 

\item The 60 GeV beam is very well tuned for the first oscillation maximum and consequently 
has a large effect due to muon neutrino disappearance.  Almost 75\% of the total muon neutrino 
events are calculated to disappear.   This factor is smaller for the 8 GeV beam because of multiple 
oscillation nodes. 

\item The neutrino contamination in the antineutrino beam is large $\sim 26\%$ for the 60 GeV beam, but
it is much smaller for the 8 GeV beam.  Nevertheless, the event rate for the antineutrino running in the 8 GeV 
beam is much more suppressed compared to the neutrino running.  This difference can be traced back to 
both the production rate of $\pi^-$ and the neutrino/antineutrino cross sections at low energies. 
The large suppression of antineutrino rates at low energies is the additional reason for the 
 complementarity between high and 
low energy running for LBNE.

\end{itemize}

\begin{table} 
\begin{tabular}{|l|c|c||c|c|} 
\hline   
Event type  &   8 GeV & 60 GeV & 8 GeV & 60 GeV   \\ 
     &  $\nu$    & $\nu$ & $\bar\nu$ & $\bar\nu$   \\
\hline  
$\nu_\mu$CC &  8900      & 66000   & 200    &  5870    \\
 w/osc          & 4600   &  17100   & 101   &    2900    \\
\hline   
$\bar\nu_\mu$CC & 76   & 1800     & 1900   &   22000      \\
 w/osc          & 36    & 850    & 1030    &  5800       \\
\hline   
$\nu_e$CC &   95         & 580    & 1    &  69      \\
$\bar\nu_e$CC & 1      & 14    & 23   &  172       \\
\hline  
\end{tabular} 
\caption{Total rate of events for the 8 GeV and 60 GeV neutrino and antineutrino beams.
The 8 GeV (60 GeV) beam is assumed to have power of 3 MW (2MW). The running time is one year 
and the detector fiducial mass is 200 kTon at 1300 km from Fermilab. The event rate after 
the disappearance of muon neutrinos by oscillations is also shown for the muon neutrino and antineutrino 
components.   } 
\label{ratetab} 
\end{table}

\section{Electron neutrino appearance and precision CP violation measurements} 

We have tabulated the performance of the water Cherenkov detector using data analysis from Super-Kamiokande (SK). 
This performance has been verified by several independent checks on SK data. The performance of 
the 200 kTon LBNE water detector is expected to be similar or better because of higher pixelation and 
much better time resolution of the photo-multiplier tubes.  Work continues to fully simulate and reconstruct 
events in the 200 kTon water Cherenkov detector. A recent eye-scan of the Monte Carlo events indicates considerable 
room for improvement beyond the Super-Kamiokande based performance by including low ring-multiplicity 
events \cite{sanjib}. 
Nevertheless, in the following we have used the 
SK based signal efficiency and background rejection since it is the most  conservative estimate.  

The calculation is performed by first using the total charged current and neutral current event spectrum 
as a function of neutrino energy and using the efficiency for selecting events reconstructed as single 
electromagnetic showers with no signatures of additional particles including decays of muons that might be 
below Cherenkov threshold. This calculation includes the expected energy resolution 
and smearing of both charged current and neutral current events.   After obtaining the spectrum of these 
events using tabulated efficiency factors, we use a tabulated  event identification likelihood efficiency (LL) 
as a function of 
{\it reconstructed} energy.  This likelihood performance 
was also tabulated using SuperK Monte Carlo that has been tuned 
to atmospheric neutrino data.    The likelihood performance tabulation 
includes the efficiency for charged current electron neutrino 
signal events and both neutral current and charged current muon neutrino events.  
The likelihood performance can be tuned to obtain high signal efficiency (for example, from  80\%LL to 40\%LL) 
and different levels of background rejection\cite{pwgrep}.  

\begin{figure} 
\mbox{
\includegraphics[angle=0,width=0.45\textwidth]{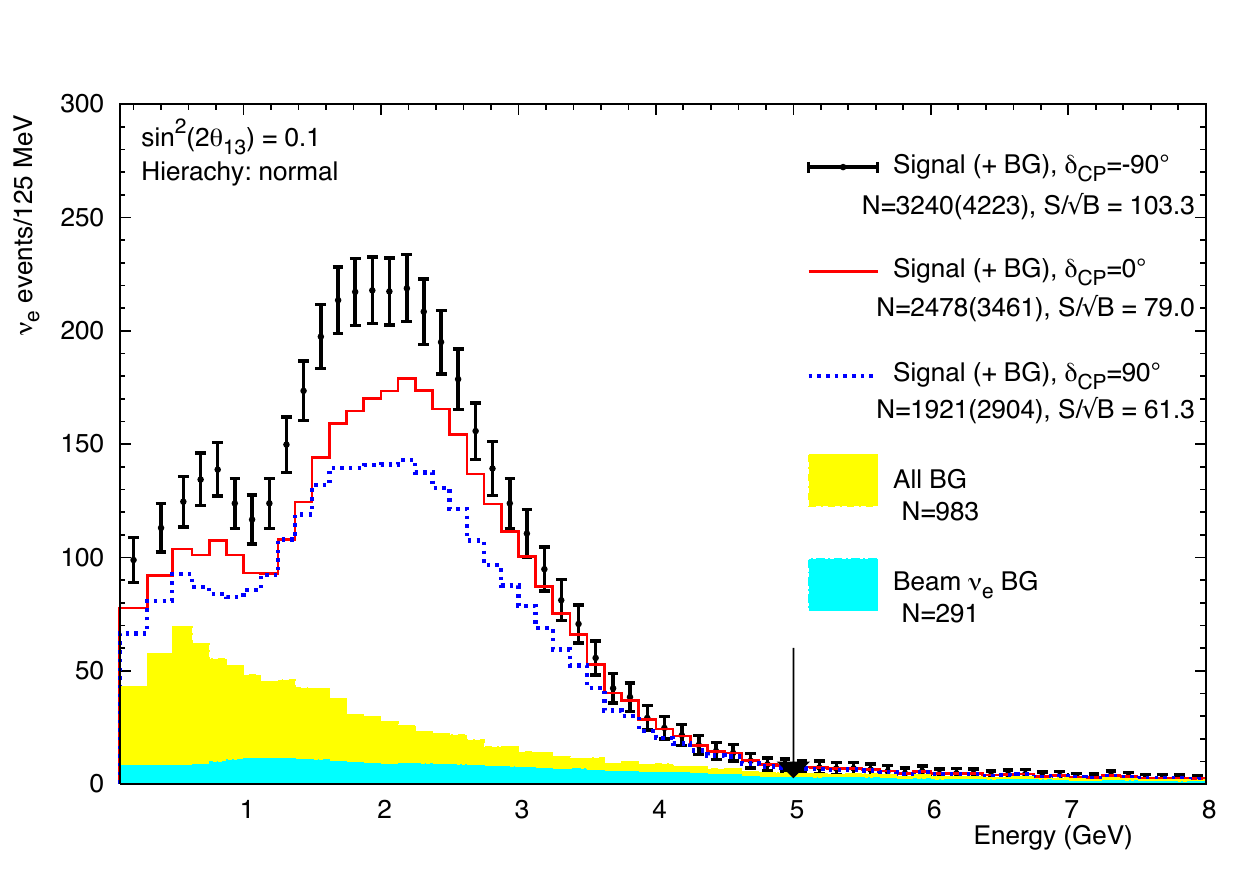}
\includegraphics[angle=0,width=0.45\textwidth]{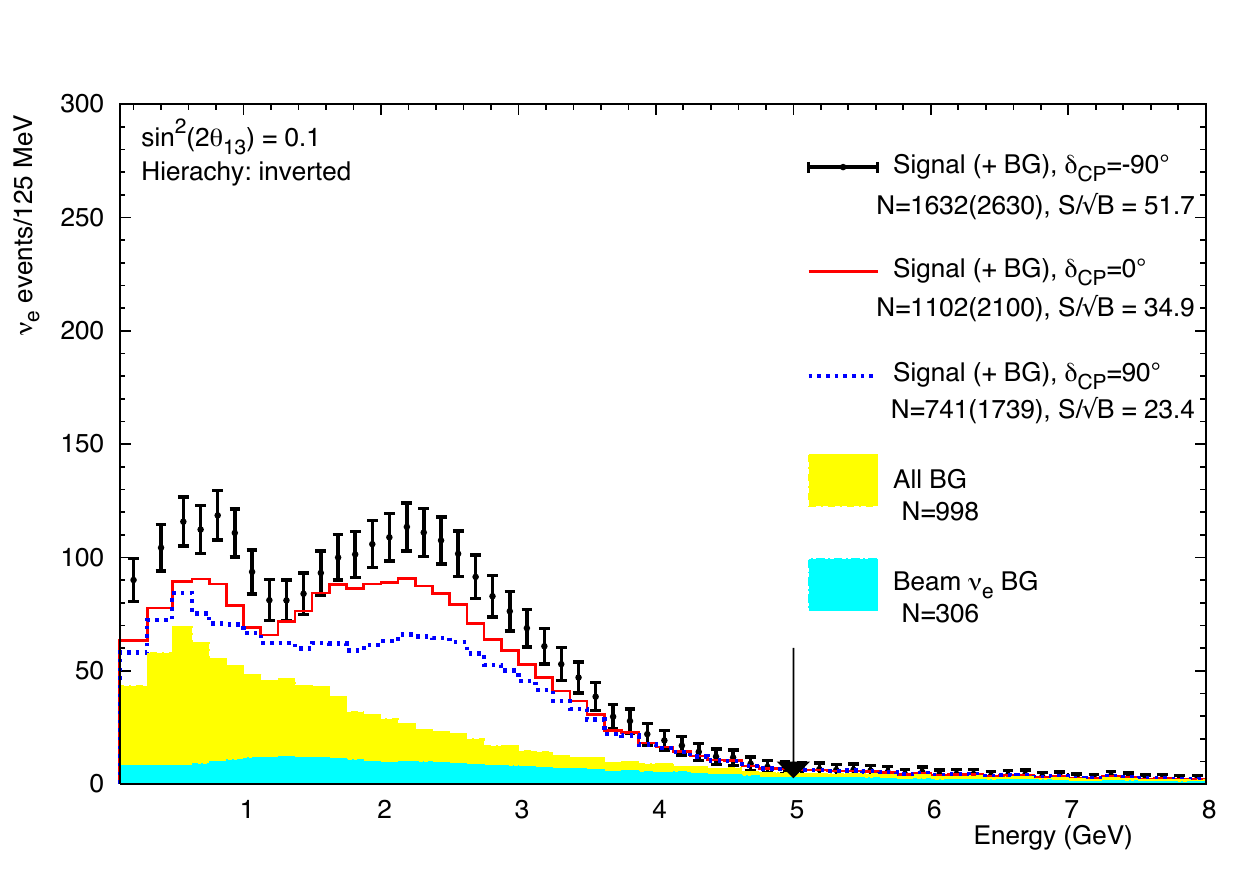}}
\mbox{
\includegraphics[angle=0,width=0.45\textwidth]{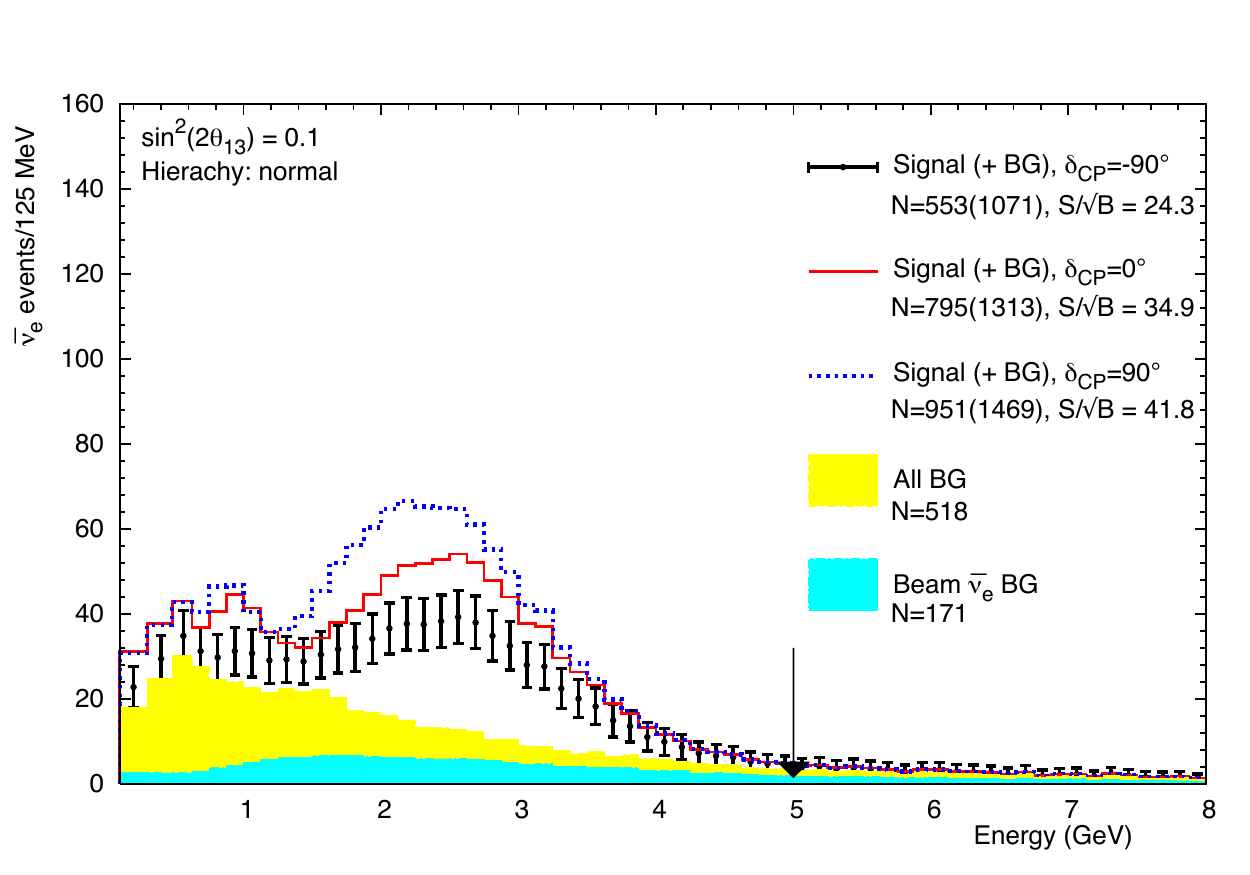}
\includegraphics[angle=0,width=0.45\textwidth]{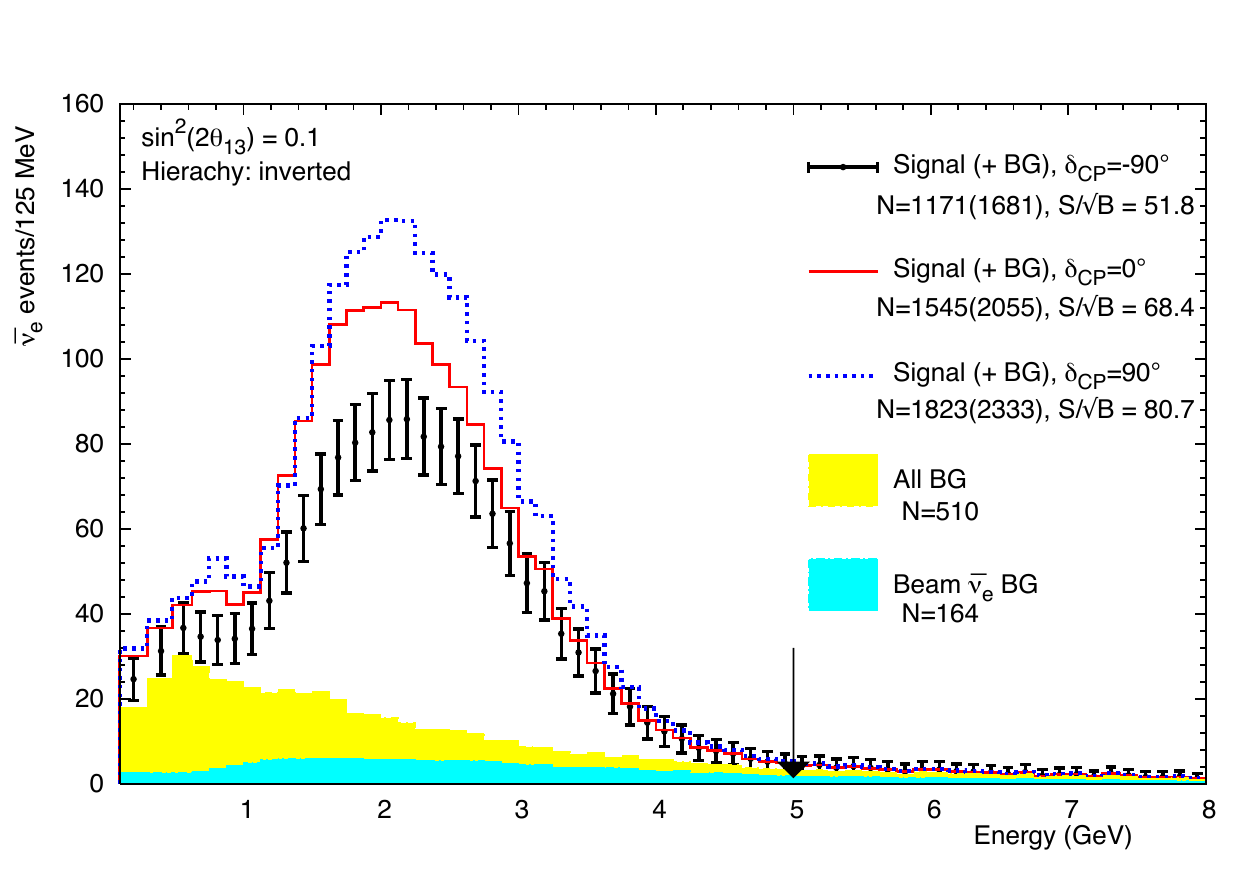}}
\caption{Event spectra using a 200 kTon water Cherenkov detector at Homestake for a neutrino beam from FNAL. 
We have assumed 2 MW of power at 60 GeV for 5 years of running for neutrinos (top plots)  and antineutrino modes 
(bottom plots).  The left plots are for normal mass hierarchy and right plots are for inverted mass hierarchy. 
We have put 
 $\sin^2 2 \theta_{13} = 0.1 $.  The curve with error bars in each plot is for $\delta_{CP} = -\pi/2$ and the 
red and blue curves are for $\delta_{CP} = 0, \pi/2 $, respectively.    The integral numbers of events are shown in 
the legends. The integral was computed up to the arrow  in the figure. 
 }
\label{gspec60} 
\end{figure}

\begin{figure} 
\mbox{
\includegraphics[angle=0,width=0.45\textwidth]{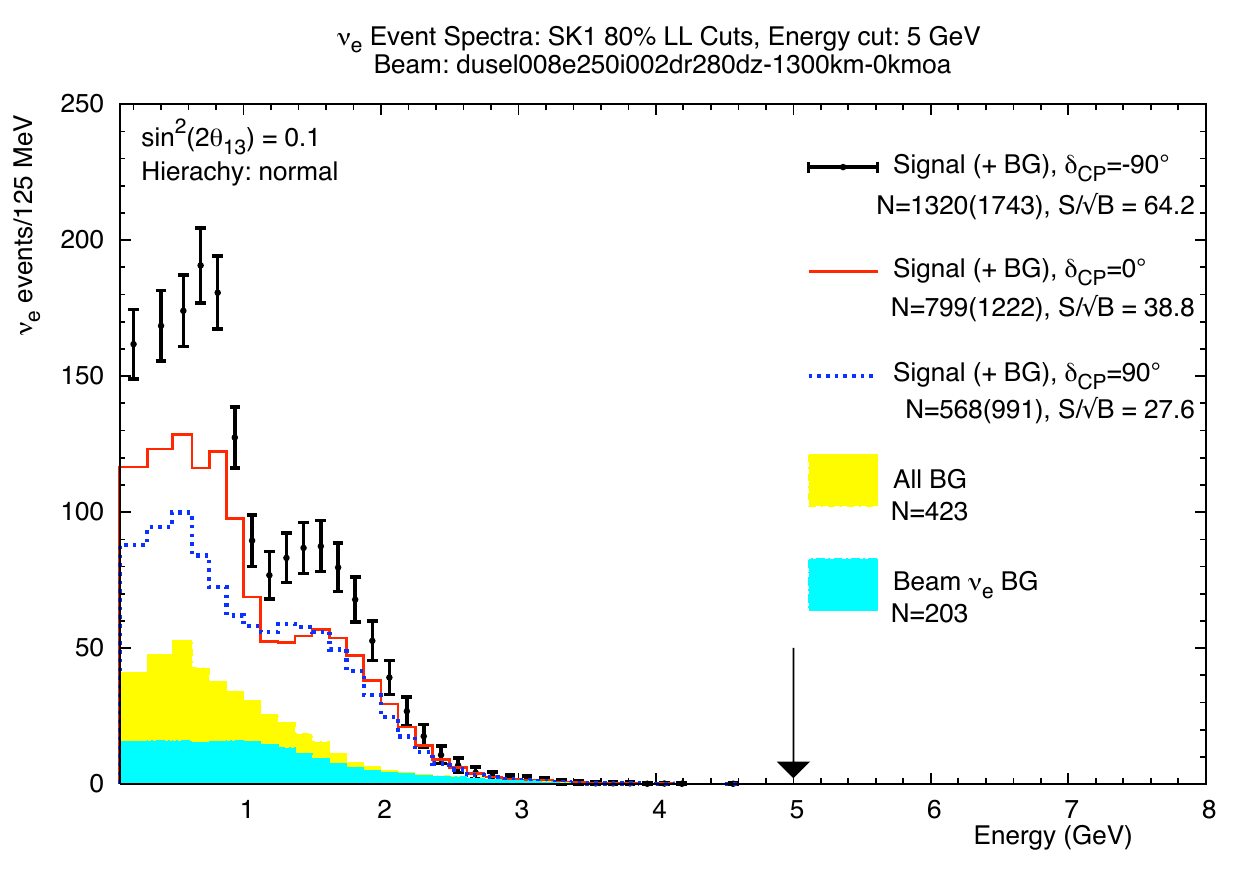}
\includegraphics[angle=0,width=0.45\textwidth]{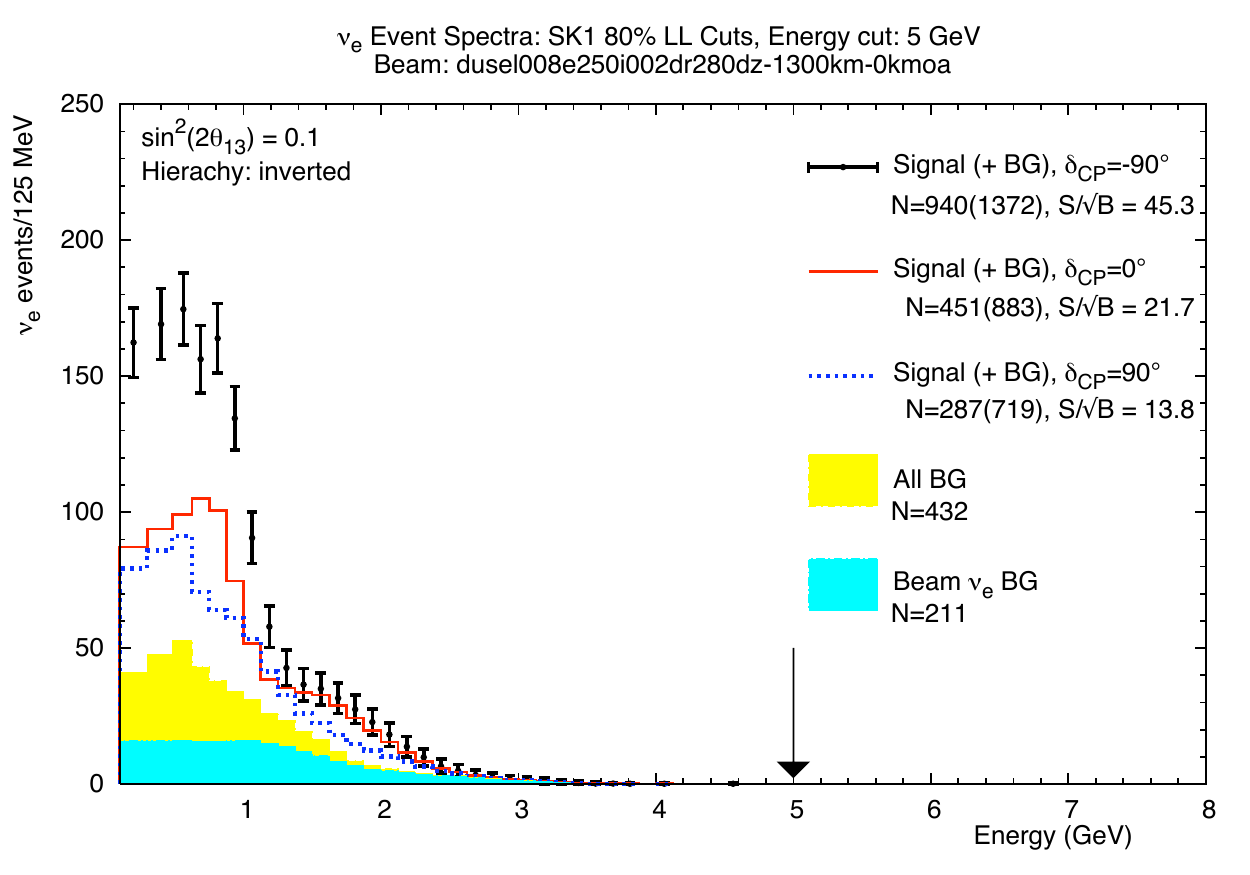}}
\mbox{
\includegraphics[angle=0,width=0.45\textwidth]{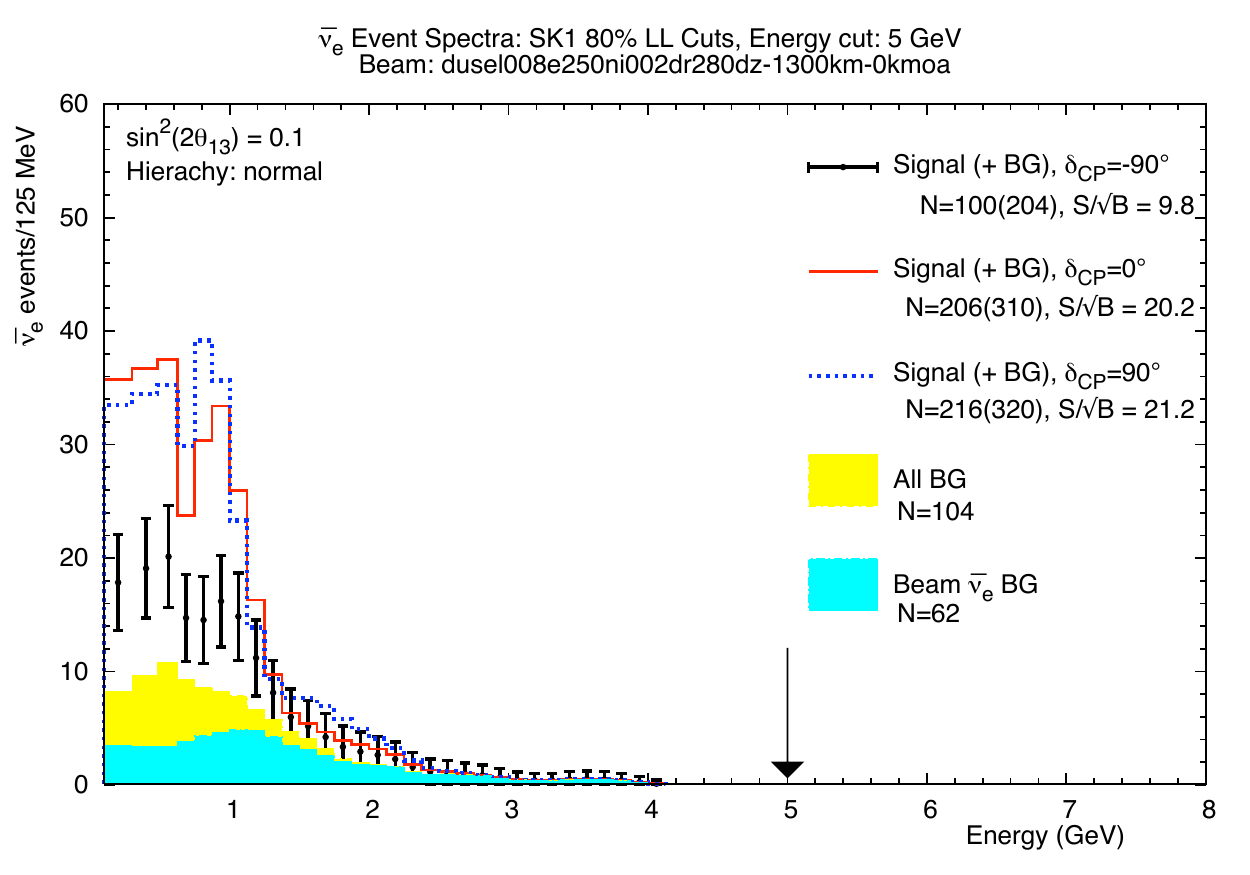}
\includegraphics[angle=0,width=0.45\textwidth]{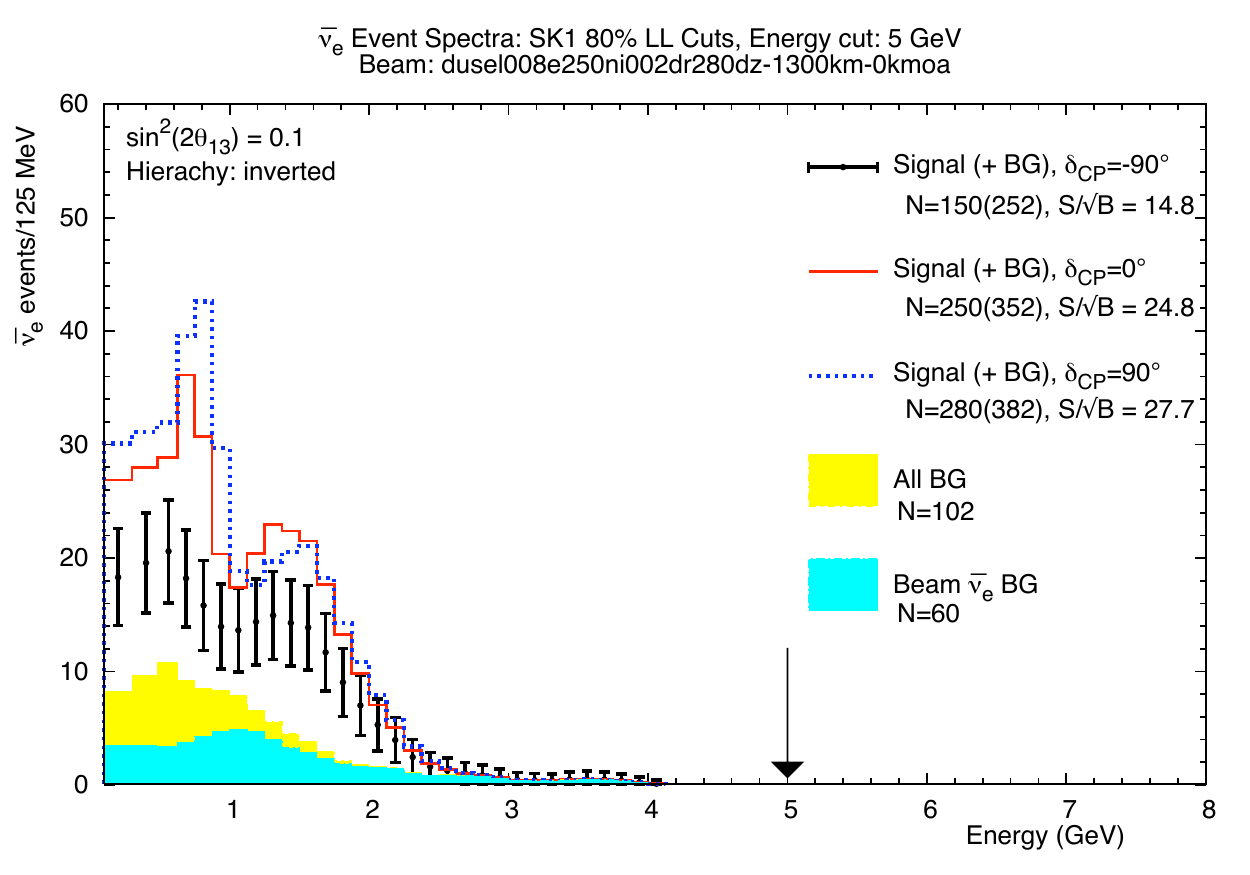}}
\caption{
Event spectra using a 200 kTon water Cherenkov detector at Homestake for a neutrino beam from FNAL. 
We have assumed 3 MW of power at 8 GeV for 5 years of running for neutrinos (top plots)  and antineutrino modes 
(bottom plots).  The left plots are for normal mass hierarchy and right plots are for inverted mass hierarchy. 
We have put $\sin^2 2 \theta_{13} = 0.1 $.  
The curve with error bars in each plot is for $\delta_{CP} = -\pi/2$ and the 
red and blue curves are for $\delta_{CP} = 0, \pi/2 $, respectively.   The integral numbers of events are shown in 
the legends. The integral was computed up to the arrow  in the figure.   
 }
\label{gspec8} 
\end{figure}

We have used the parameterized performance for the water detector using the 60 GeV and 8 GeV beams to calculate the 
reconstructed spectra with the proper resolution and efficiencies in figures \ref{gspec60} and \ref{gspec8}. 
For the 60 GeV running we have used a likelihood performance that has 40\% efficiency for signal in figure \ref{gspec60}. 
The lower efficiency cut reduces the neutral current background above 2 GeV to be approximately the same as the 
beam contamination of electron neutrinos and provides a spectrum  with a large signal to background ratio for 
both neutrino and antineutrino running.   

For the 8 GeV running we have used a likelihood performance that has 80\% efficiency for signal in figure \ref{gspec8}. 
Such a choice uses the advantages of running a low energy beam which will have lower backgrounds from both the
neutral currents and the beam contamination.  The advantages of the lower energy beam can be clearly seen in the 
large CP  effects below 1 GeV. The appearance signal is seen to vary by more than a factor of 2 over the 
range $\delta_{CP}  = -\pi/2 \to \pi/2$. It is clear that for the 8 GeV beam, the antineutrino event rate 
will be low. The principal reason for the low rate is the antineutrino charged current cross section at 
low energies.  In particular, the quasi-elastic cross section for antineutrinos reduces much more rapidly 
at low energies than for neutrinos. The second reason is the lower production of $\pi^-$ mesons by lower energy 
protons.  It should also be noted that the large CP effects below 1 GeV are only weakly affected by mass hierarchy. 

A simultaneous run of beams at 60 GeV and 8 GeV will provide excellent coverage of appearance spectra across 
the entire energy region. Moreover, the beam running could be managed in such a way that the two beams provide 
opposite polarity beams. For example,  the 8 GeV beam could run in the neutrino mode and the 60 GeV beam 
could run in the 
antineutrino mode. In the case of inverted hierarchy, such a mode of running could be extremely beneficial.

\begin{figure} 
\mbox{
\includegraphics[angle=0,width=0.3\textwidth]{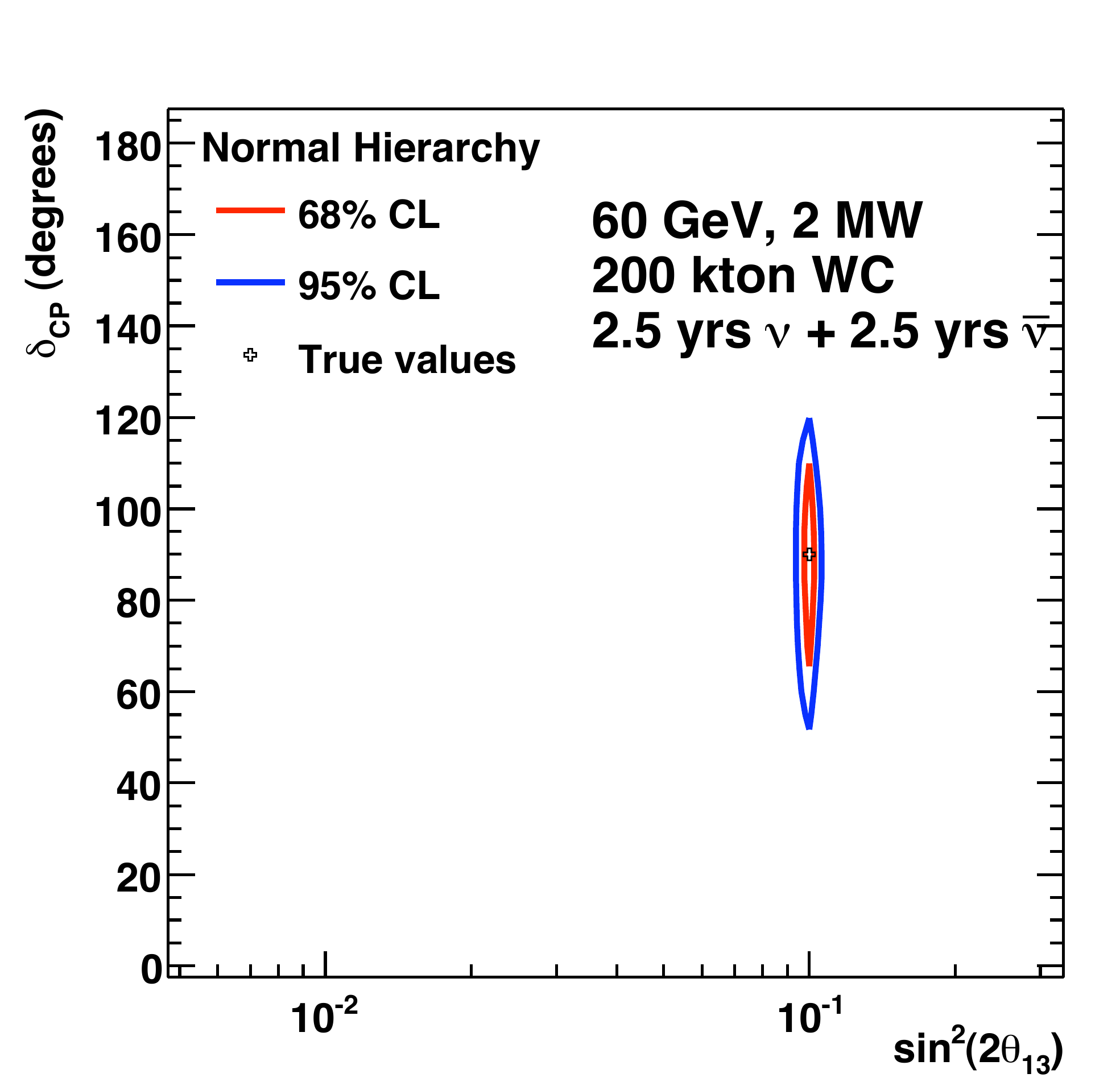}
\includegraphics[angle=0,width=0.3\textwidth]{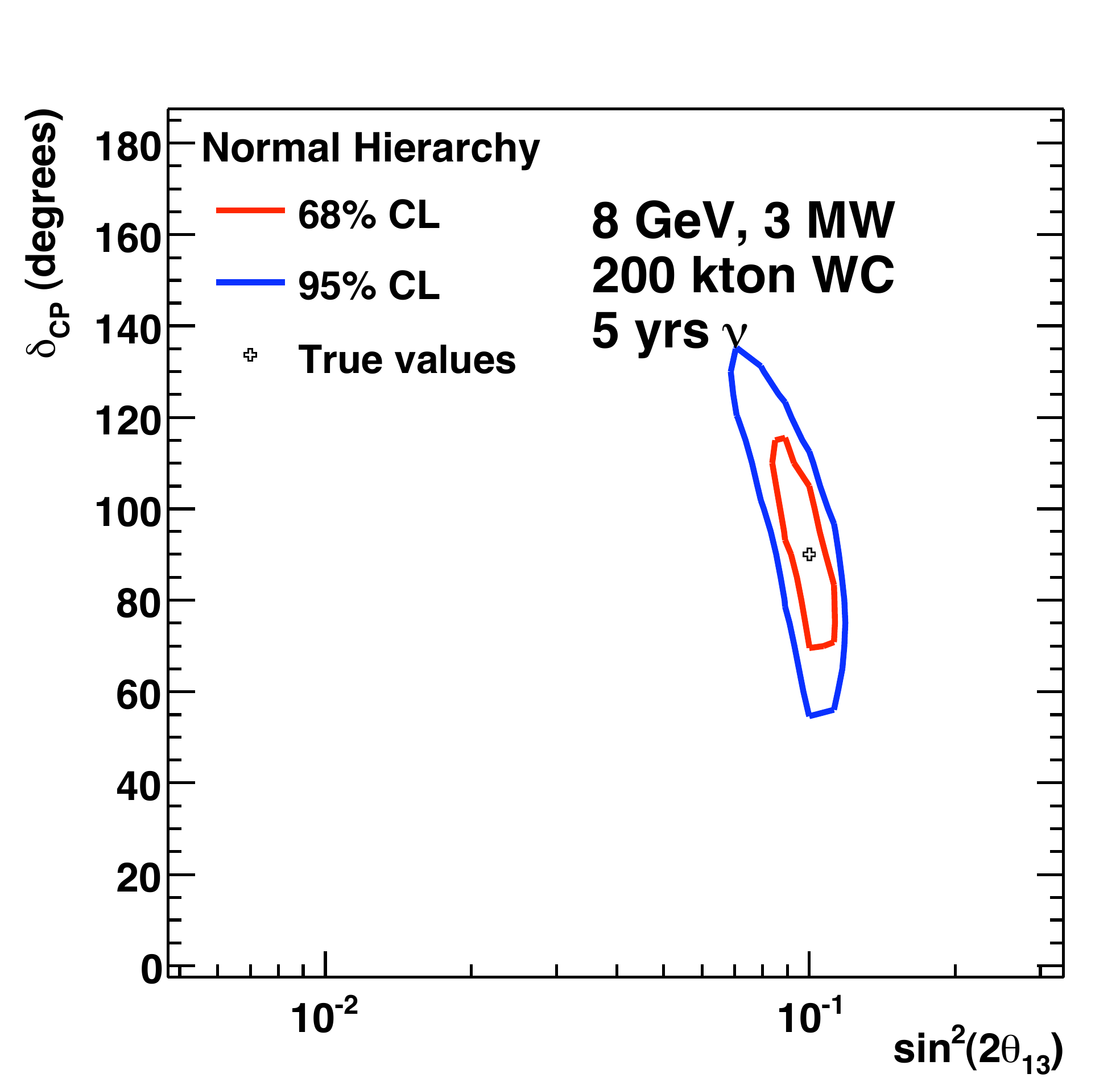} 
\includegraphics[angle=0,width=0.3\textwidth]{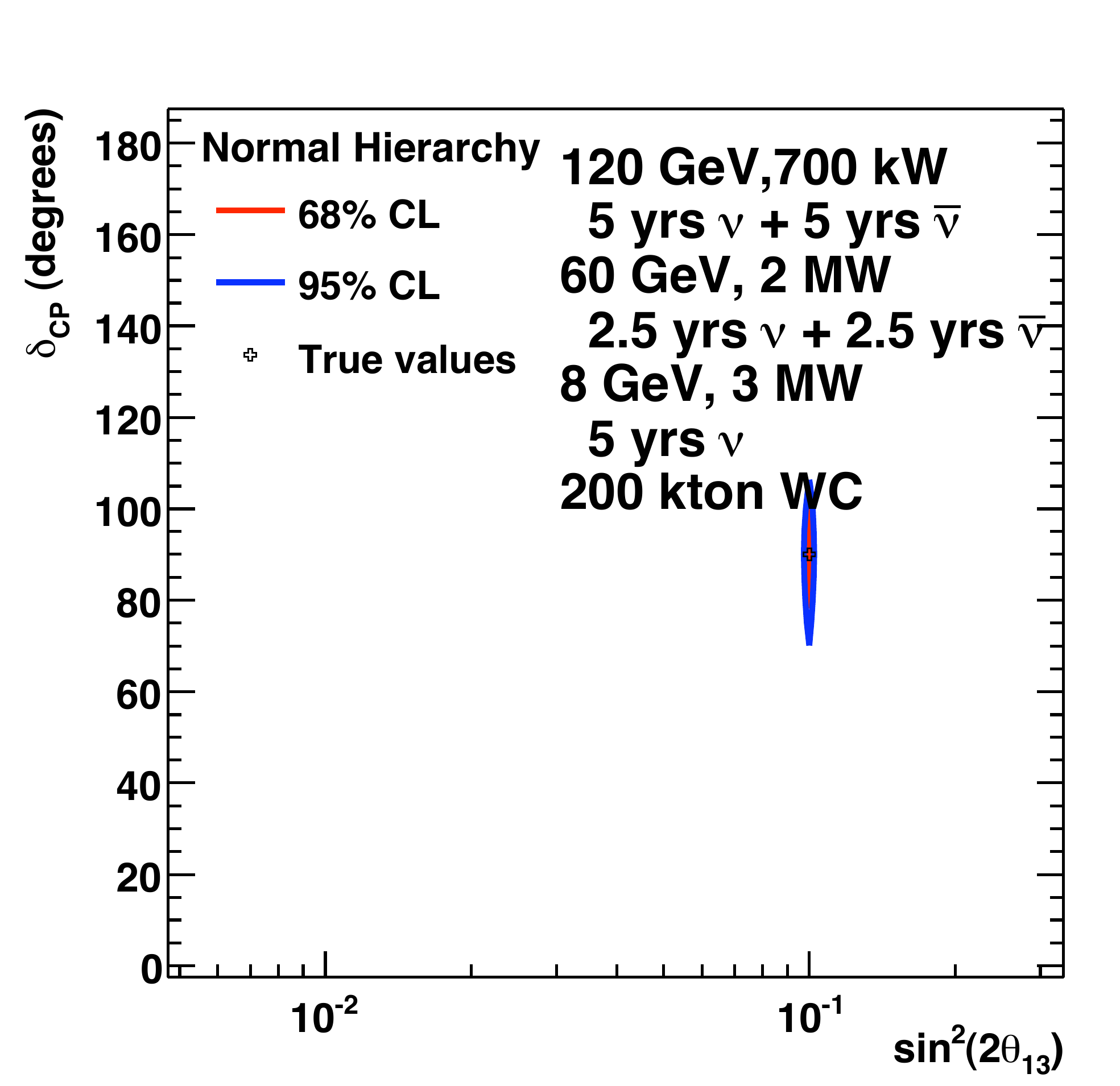}}
\caption{2 sigma and 1 sigma confidence level measurements of $\sin^2 2 \theta_{13}, \delta_{CP}$ with different running conditions. Left-most plot is for running 60 GeV beam for 2.5 yrs each in neutrino and antineutrino modes. Middle plot is 
for running the 8 GeV beam only in the neutrino mode for 5 years. The right-most plot is the combination of both 
running conditions combined with preproject-X running of 10 yrs with 120 GeV and 700 kW beam. 
For this calculation we have assumed $\sin^2 2 \theta_{13} = 0.1, \delta_{CP} = \pi/2$ and 
normal mass ordering.   }
\label{buble} 
\end{figure}

\begin{figure} 
\mbox{
\includegraphics[angle=0,width=0.3\textwidth]{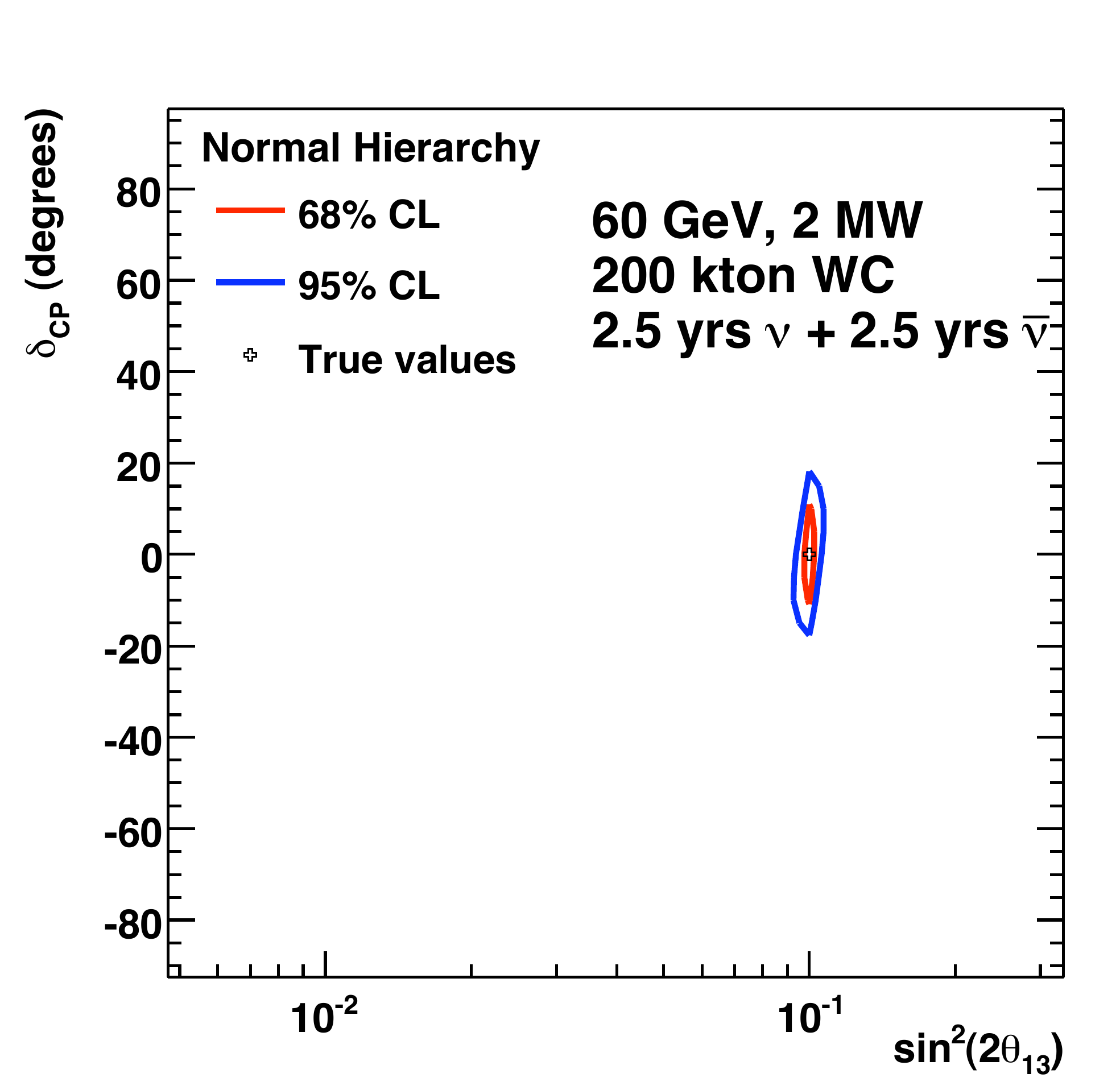}
\includegraphics[angle=0,width=0.3\textwidth]{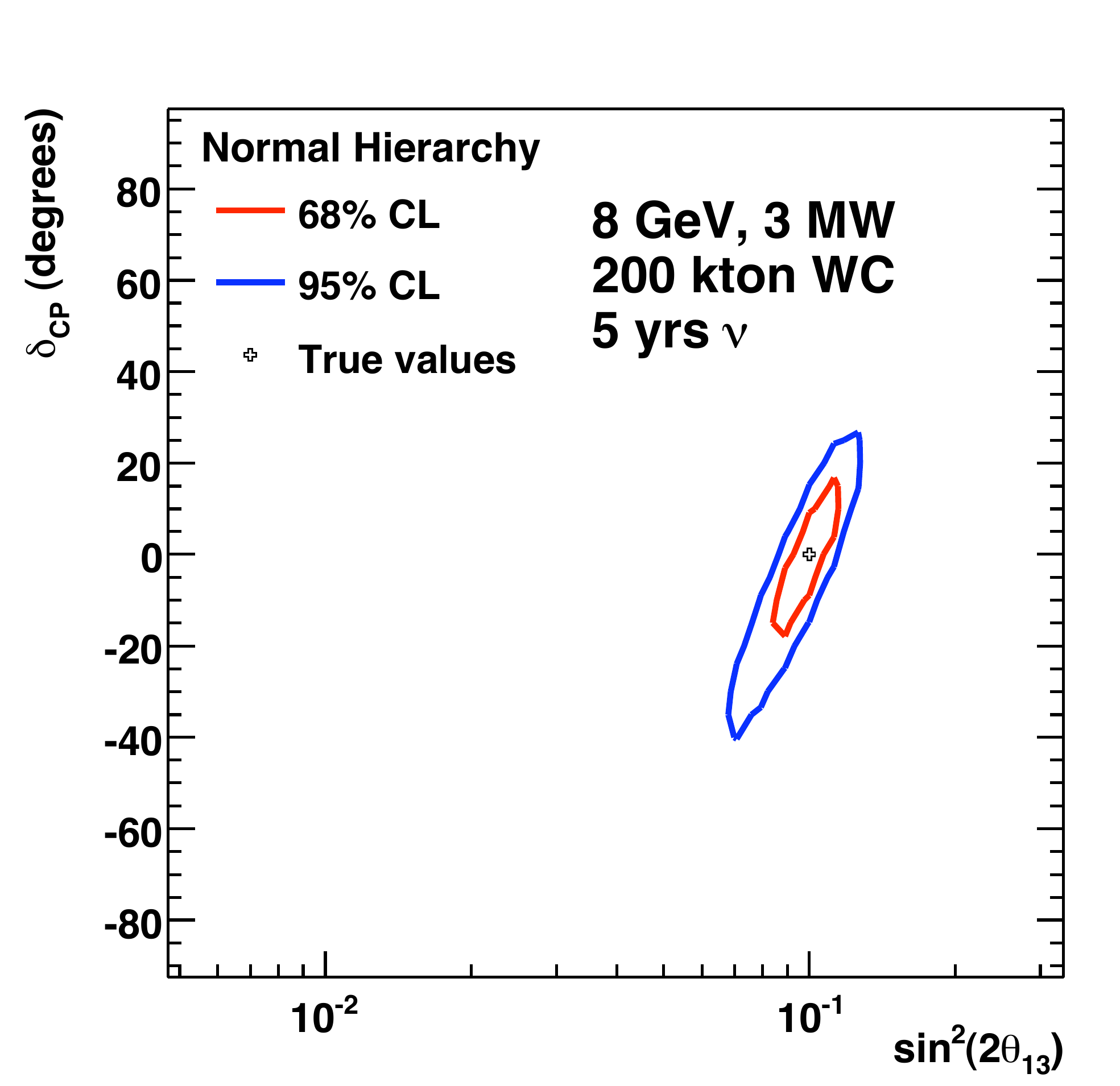} 
\includegraphics[angle=0,width=0.3\textwidth]{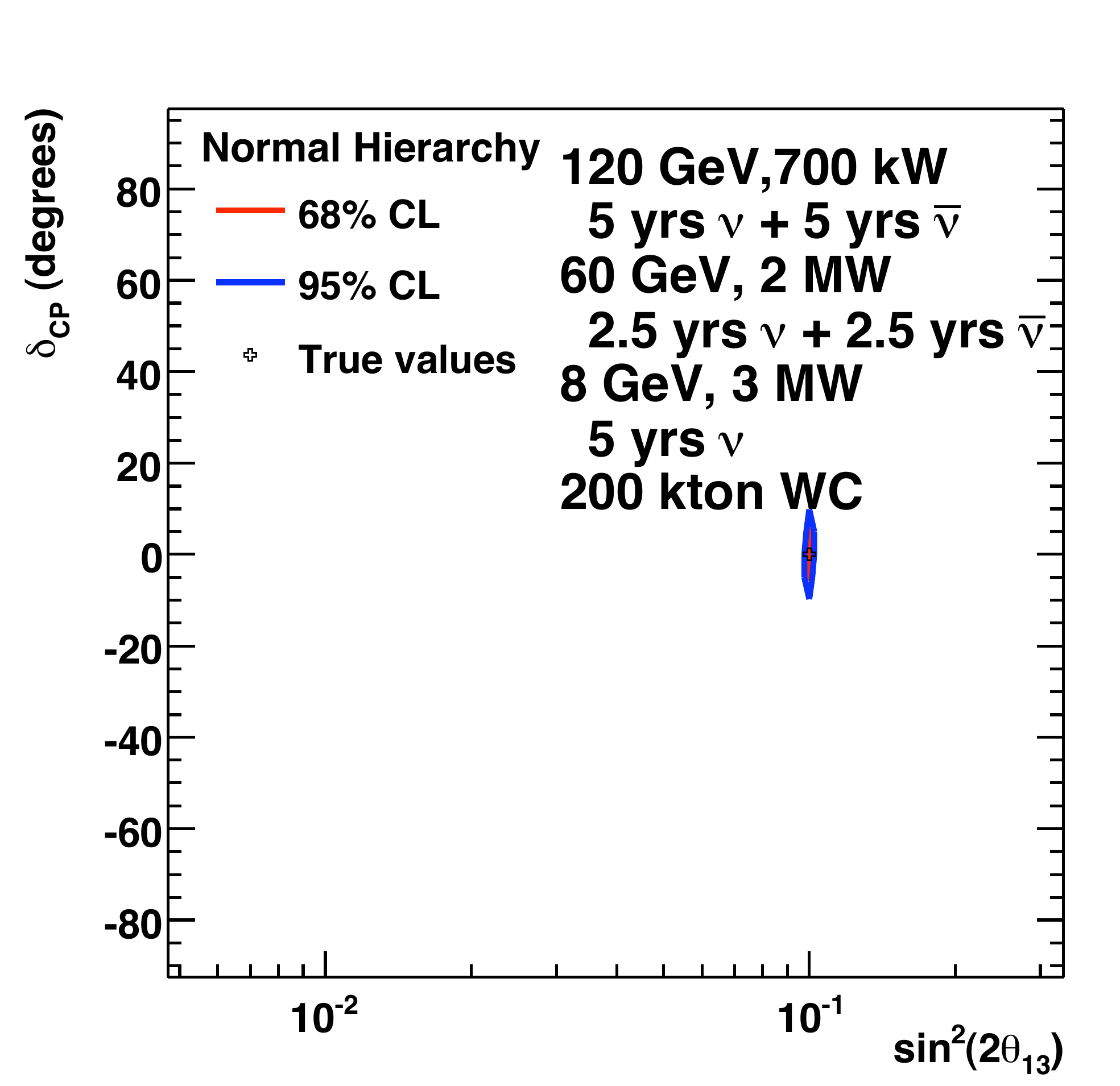}}
\caption{2 sigma and 1 sigma confidence level measurements of $\sin^2 2 \theta_{13}, \delta_{CP}$ with different running conditions. Left-most plot is for running 60 GeV beam for 2.5 yrs each in neutrino and antineutrino modes. Middle plot is 
for running the 8 GeV beam only in the neutrino mode for 5 years. The right-most plot is the combination of both 
running conditions combined with preproject-X running of 10 yrs with 120 GeV and 700 kW beam.  
For this calculation we have assumed $\sin^2 2 \theta_{13} = 0.1, \delta_{CP} = 0$ and 
normal mass ordering.   }
\label{buble0} 
\end{figure}

We have used the spectra shown in figures \ref{gspec60} and \ref{gspec8} to calculate the precision with which 
$\sin^2 2 \theta_{13}, \delta_{CP}$ could be measured.  
These calculations were performed using the GLoBES software program \cite{globes}. 
 We have investigated the sensitivity with respect to moving the  energy threshold 
From 100 MeV to 500 MeV and found the 
sensitivity to be only weakly dependent on this threshold.   
The precision is shown in figures \ref{buble} and \ref{buble0}. 
For this plot we have used a total of 5 yrs of running in 60 GeV and in 8 GeV beams 
(as remarked earlier this could be simultaneous). 
The 60 GeV running is split in neutrino 
and antineutrino modes, but the 8 GeV running is in only the neutrino mode.
For the combined plot, we have also included the data from the pre-project-X period for LBNE
which would span a total of 10 years with 700 kW of running at 120 GeV \cite{pwgrep}. 
The precision is dominated by the 60 GeV running after Project-X, nevertheless the pre-Project-X running 
certainly helps.   

\begin{figure} 
\mbox{
\includegraphics[angle=0,width=0.45\textwidth]{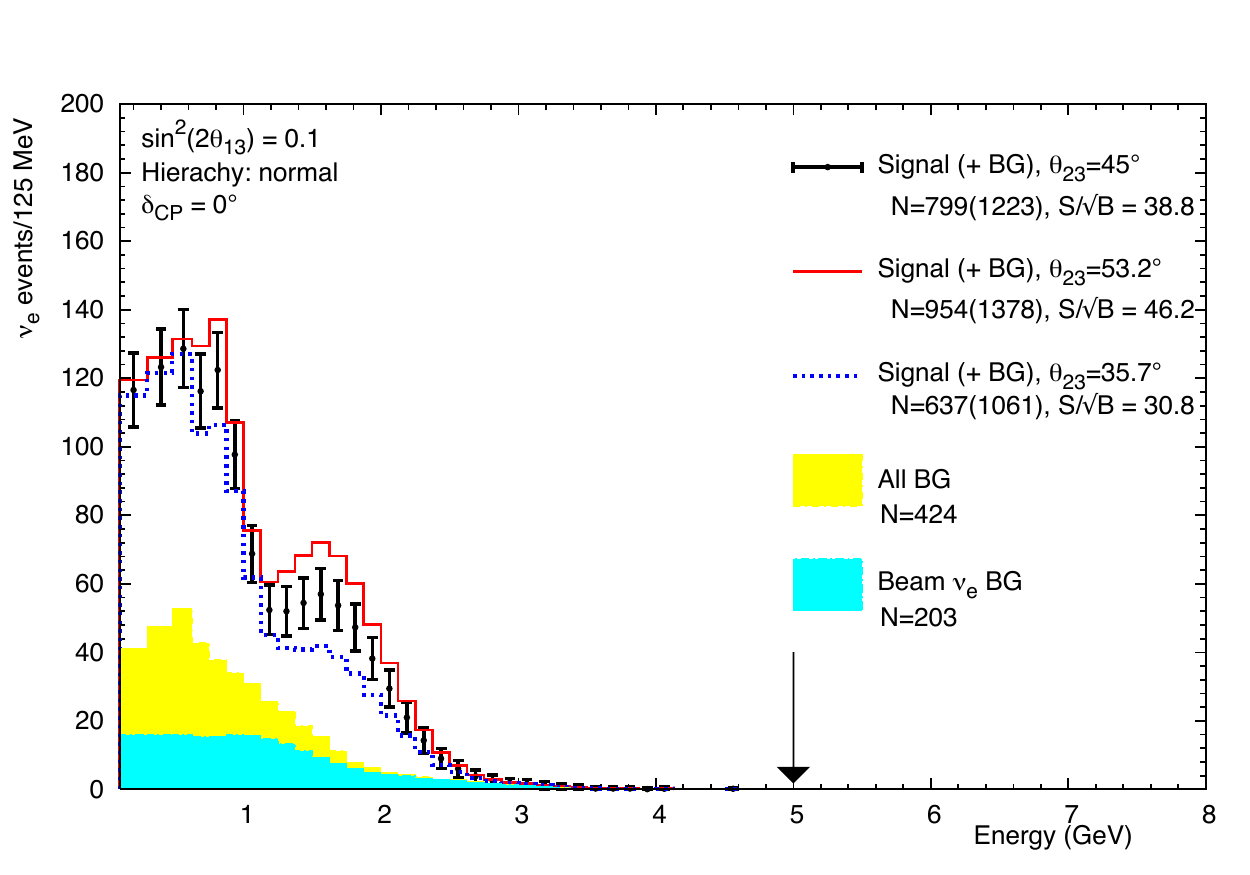}
\includegraphics[angle=0,width=0.45\textwidth]{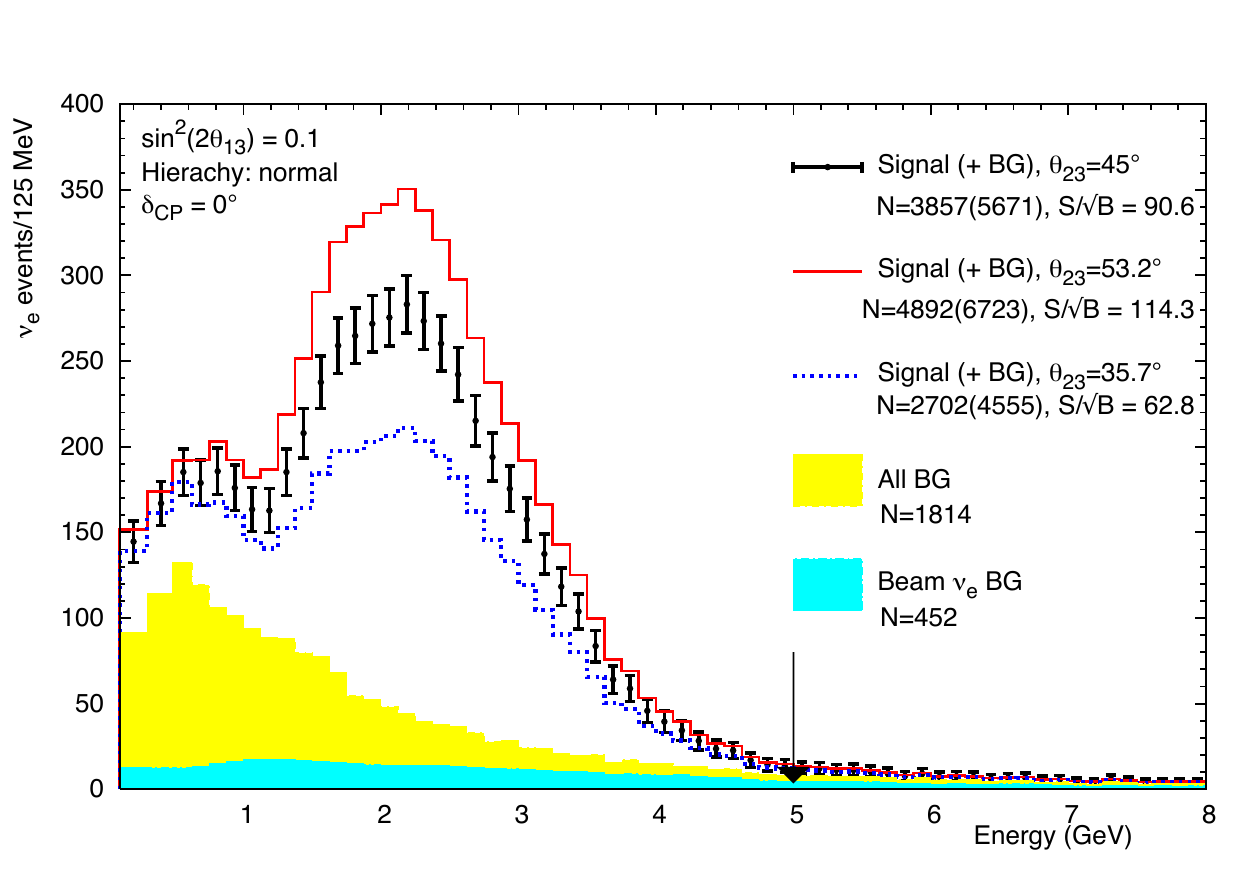}}
\caption{Effect of $\theta_{23}$  variation on the appearance spectra from 8 GeV running (left)  
and 60 GeV running (right). 
 The integral numbers of events are shown in 
the legends. The integral was computed upto the arrow  in the figure. 
 }
\label{th23exam} 
\end{figure}

 We will remark on a few features of this 
measurement: 

\begin{itemize} 
\item The two beams cover very different energy regions and have  independent sensitivity. The 60 GeV 
data will be affected by large matter effects and will result in the resolution of the mass hierarchy with 
very high ($>15 \sigma$) confidence. Once the mass hierarchy is known, the same data can be used for additional 
constraints on the matter potential or other contributions to the oscillation as described below.  

\item The 60 GeV data provides high precision on $\sin^2 2 \theta_{13}$ ($\sim$ few percent). 
The 8 GeV data with neutrino only 
running will have high precision to $\delta_{CP}$, but it will be correlated to $\theta_{13}$. When the two sets of 
data are combined, a measurement of $\delta_{CP}$ with an error of $\pm 10^\circ$ ($\pm 5^\circ$) at 
$\delta_{CP} = \pi/2$ ($\delta_{CP} = 0$)  is possible.  

\item It is certainly possible to run the 60 GeV beam for much longer  running time and obtain approximately the 
same precision on ($\theta_{13}, \delta_{CP}$) as the combined 8 GeV and 60 GeV result. However, this will not 
provide the same independent constraint on the parameters that could lead to discovery of new phases or 
interactions. 

\item The figures \ref{buble} and \ref{buble0}  were computed for normal mass hierarchy. 
For the inverted mass hierachy 
the sensitivity is the same because the statistical merit of the combined neutrino and 
antineutrino data set for the 60 GeV beam 
remains approximately the same. For the inverted hierarchy there will be higher signal in the
 antineutrino mode and lower in the neutrino  mode. 
However, it should be noted that for the inverted hierarchy the 8 GeV beam should still be  run in 
the neutrino mode.

\item The two independent data sets form independent constraints on the neutrino oscillation parameters. If there is 
any additional potential difference between the two mass eigenstates due to new physics it will become evident 
as a shift in the measured parameters from these two datasets at different energies. Such a shift could appear 
even if there is no explicit CP violation in the 3-generation picture as represented by the parameter $\delta_{CP}$. 
This measurement precision could be used as a model independent parameter to test for new physics.

\item  All calculations of sensitivity so far have assumed that the true value for $\sin^2 2 \theta_{23} = 1.0$ or 
$\theta_{23} = \pi/4$. 
However, $\theta_{23}$ is actually not very well known. The measurement ranges from $35.7^\circ$ to $53.2^\circ$ at 3$\sigma$. 
In particular, the value could be above or below $45^\circ$.  The deviation of $\theta_{23}$ from  maximum mixing is of 
high interest to GUT theorist who can predict this particular mixing angle through various models. The deviation of 
$\theta_{23}$ from $45^\circ$ might be indicative of the relationship between quark and lepton mixing.   

Our current calculations include a precise determination of $\sin^2 2 \theta_{23}$ using the disappearance data from
LBNE. This internal constraint is included in the sensitivities published  for LBNE.  However, the disappearance data
is not sensitive to the octant of $\theta_{23}$.  The appearance data has excellent sensitivity to the angle 
$\theta_{23}$ as shown in equation \ref{qe1}. The first term of the probability depends on $\sin^2 \theta_{23}$. 
This can be seen in figure \ref{th23exam}. The oscillation probability at high energies using the 60 GeV beam 
is seen to 
have a large dependence on $\theta_{23}$, but the 8 GeV spectra do not display the same sensitivity. 
A joint fit using 8 GeV and 60 GeV data is expected to resolve the octant of $\theta_{23}$.  We will display this 
calculation in an update to this note.

\end{itemize}

\section*{Conclusion}

We have examined a method to obtain multiple constraints in the neutrino oscillation sector 
using Project-X with    high statistics, low energy neutrinos over long distances. This is motivated by 
the recent indications of a modest value for $\theta_{13} \sim 9^\circ$.  
For oscillation physics at low energies the charged current cross section is dominated by quasielastic scattering.
For the quasielastic final state any detector capable of measuring single lepton final states is adequate, however 
as a result of the work on LBNE 
the water Cherenkov detector has been shown  to be cabable of achieving the needed target mass to obtain the  
statistical precision. Furthermore, LBNE related simulation and reconstruction work has resulted in 
good understanding of the performance of this detector in the beams that we have described above.  
 And therefore we have used a 200 kTon detector to obtain event spectra and sensitivity in 
the ($\theta_{13}, \delta_{CP}$) space.  

Another strategy to obtain events at high values of L/E is by utilizing a much longer baseline such as 
2500 to 2700 km from FNAL to California or Oregon. We suggest a renewed examination of such a strategy 
and optimization of the beam spectrum for much longer baselines.  For example, running with different tunes of 
the FNAL neutrino could allow good coverage of L/E over much longer baselines.

Project-X will allow great flexibility in proton beam energy and power that can be delivered for neutrino 
beams. In particular, we find the possibility of an upgrade to the pulsed LINAC for Project-X to deliver high 
power at 8 GeV very interesting.  With such an upgrade 
high power 8 GeV running can be simultaneous with  60 GeV running because 
only   266  kW of the 8 GeV  beam will be fed to the Main Injector to make 
2 MW of 60 GeV.   The rest of the almost 4 MW of 8 GeV could be used separately for low energy neutrino production. 
A combination of 8 GeV and 60 GeV running will allow coverage of multiple oscillation nodes with high statistics. 

The combined data set from 8 GeV and 60 GeV running will allow multiple independent constraints on the mixing 
parameters. The shift  in the $\theta_{13}, \delta_{CP}$ solutions from these data sets could  be used as a  model 
independent parameter to test for new physics.  If the data sets match, very high precision ranging from $\pm 5^\circ$ to 
$\pm 10^\circ$  on $\delta_{CP}$ is possible with a few years of running.

\pagebreak


\begin{thebibliography}{99}
\bibitem{stevebob} The proton beam intensity and duty factor were provided by Steve Holmes 
and Robert Tschirhart.
\end{thebibliography}
\end{document}